\documentstyle[12pt]{article}
\def\NPB{{\it Nucl. Phys. }{\bf B}}
\def\PL{{\it Phys. Lett. }}
\def\PRL{{\it Phys. Rev. Lett. }}
\def\PRD{{\it Phys. Rev. }{\bf D}}
\def\CQG{{\it Class. Quantum Grav. }}

\def\IJMPA{{\it Int. J. Mod. Phys. }{\bf A}}
\def\MPL{{\it Mod. Phys. Lett. }}

\def\ie{{\it i.e.},}

\newcommand{\eq}{\begin{equation}}
\newcommand{\en}{\end{equation}}
\newcommand{\eqn}{\begin{eqnarray}}
\newcommand{\enn}{\end{eqnarray}}
\newcommand{\nn}{\nonumber }
\newcommand{\beq}{\begin{equation}}
\newcommand{\eeq}{\end{equation}}

\let\a=\alpha
\let\b=\beta

\let\b=\beta
\def\cC{{\cal C}}
\def\const{\hbox{\it const.\/}}
\def\CP#1{\relax\ifmmode\IP^{#1}\else\IP$^{#1}$\fi}
\def\rd{{\rm d}}
\let\vd=\partial
\def\define{\buildrel{\rm def}\over=}
\let\f=\phi
\def\enh{enhan\c{c}on}

\def\cE{{\cal E}}
\let\F=\Phi
\let\G=\Gamma
\def\Imm{\Im m}
\def\inv#1{{\textstyle{1\over#1}}}
\def\IP{\relax\leavevmode{\rm I\kern-.18em P}}
\def\cL{{\cal L}}
\let\l=\lambda
\def\Ione{\relax\leavevmode{\rm 1\kern-3pt l}}
\let\q=\theta
\let\Q=\Theta
\let\p=\pi

\def\re{r\mkern-2mu_{\rm e}}
\def\ri{r\mkern-1mu_{\rm i}}
\def\rF{r_\Phi}
\def\rs{r\mkern-2mu_{\rm s}}
\def\ro{r\mkern-2mu_{\rm o}}
\def\Ree{\Re e}
\let\t=\tau
\def\cT{{\cal T}}
\let\To=\Rightarrow
\def\cV{{\cal V}}

\let\w=\omega

\def\IR{\relax\leavevmode{\rm I\kern-.18em R}}
\def\ZZ{\relax\leavevmode
              \ifmmode\mathchoice
              {\hbox{\sf Z\kern-.4em Z}}
              {\hbox{\sf Z\kern-.4em Z}}
              {\lower.9pt\hbox{\scriptsize\sf Z\kern-.36em Z}}
              {\lower1.2pt\hbox{\tiny\sf Z\kern-.36em Z}}
               \else{\sf Z\kern-.4em Z}\fi}
\def\RR{\relax\leavevmode
              \ifmmode\mathchoice
              {\hbox{\sf R\kern-.4em R}}
              {\hbox{\sf R\kern-.4em R}}
              {\lower.9pt\hbox{\scriptsize\sf R\kern-.36em R}}
              {\lower1.2pt\hbox{\tiny\sf R\kern-.36em R}}
               \else{\sf R\kern-.4em R}\fi}

\catcode`@=11   % WATCH OUT !!!!! This must be reset in the end !!!!!
\def\resetby#1#2{\@addtoreset{#2}{#1}}
\def\seceq{\@addtoreset{equation}{section}% Numbers Eq.s within Sect.s
        \def\theequation{\thesection.\arabic{equation}}} % (Sect.Eq)
\catcode`@=12                   % You see ?

\def\Label#1{\label{#1}%
    \smash{\hbox to0pt{\raise1ex\hbox{\tiny[#1]}\hss}}}
\def\noLabels{\let\Label=\label}

\thicklines     

\seceq
\noLabels % uncomment for final production
\begin{document}
\input epsf.tex

\begin{titlepage}
\begin{flushright}
CITUSC/00-061 \\
hep-th/0012042\\
\end{flushright}

\begin{center}

%{\Large\bf PROBING NON-SUPERSYMMETRIC STRING VACUA} \\[10mm]
    %
{\Large\bf PROBING NAKED SINGULARITIES IN\\[5mm]
              NON-SUPERSYMMETRIC STRING VACUA} \\[10mm]
{\bf P. Berglund\footnote{e-mail: berglund@citusc.usc.edu} } \\[1mm]
        CIT-USC Center for Theoretical Physics\\
       Department of Physics and Astronomy\\
       University of Southern California\\
       Los Angeles, CA 90089-0484\\[5mm]
{\bf T. H\"{u}bsch\footnote{e-mail: thubsch@howard.edu}%
       $^,$\footnote{On leave from the ``Rudjer Bo\v skovi\'c'' Institute,
                 Zagreb, Croatia.} } \\[1mm]
       Department of Physics and Astronomy\\
       Howard University\\
       Washington, DC 20059\\[5mm]
{\bf D. Minic\footnote{e-mail: minic@physics.usc.edu} } \\[1mm]
       CIT-USC Center for Theoretical Physics\\
       Department of Physics and Astronomy\\
       University of Southern California\\
       Los Angeles, CA 90089-0484\\[10mm]
{\bf ABSTRACT}\\[3mm]
\parbox{5in}{
We present a detailed analysis of
non-supersymmetric spacetime varying string vacua
which can lead to an exponential hierarchy between the
electroweak and the gravitational scales. In particular, we identify
a limit in which these vacua can be interpreted as supersymmetric
vacua of F-theory.
Furthermore, we study the properties of these solutions as seen by
$D7$-brane probes and establish
a non-supersymmetric analogue of the
\enh\ mechanism.}
\end{center}

\end{titlepage}

\section{Introduction}

In a recent paper~\cite{bhmone}
we considered $p{+}1$-dimensional cosmic defects
(and in particular $p=D{-}3$)
embedded in a $D$-dimensional spacetime\footnote{
The possibility that our $3{+}1$-dimensional world can be viewed as a
cosmic defect (brane) in a higher-dimensional theory has been studied by
many authors, for example~\cite{rubakov, visser, gell,
anton, georgia, horava, braneTH, savas, branew1, branew2, branew3,
rs1, rs2}.}. It was explicitly shown
how these cosmic branes correspond to non-supersymmetric
string vacua which naturally can lead to an exponential
hierarchy between the $D$ and $D-2$ dimensional scales.
    These types of models have been
considered before in the literature~\cite{cohen, cp, RG, tony}.
The exponential hierarchy  between
the electroweak and gravitational scales arises from
non-trivial warp factors in the metric
and is naturally generated by the string coupling of $O(1)$.
(The r\^ole of warp factors in string theory and their relationship to
cosmic brane models
have been studied  in~\cite{cvetic, verlinde, greene, mayr}.)

As pointed out in~\cite{bhmone}, our solutions resemble the stringy cosmic
strings of~\cite{vafa, gh}. The general framework
is that of a higher-dimensional string theory compactified on a
Calabi-Yau (complex) $n$-fold, $M_n$, some moduli of which are
allowed to vary over part of the non-compact space. In the
uncompactified Type~IIB theory, the r\^ole of space-dependent moduli
is played by the dilaton-axion system, very much like in Vafa's
description of F-theory~\cite{FTh}. The cosmic defect (brane)
appears as a singularity of the induced spacetime metric, with its
characteristics governed by the energy momentum
tensor of the moduli.
In addition, there is a naked singularity, located
at a finite proper distance from the core of the brane, which is the
source of the exponential hierarchy.

Our solutions do not saturate the BPS bound in general, supersymmetry being
broken in the presence of the naked singularity, and can be thought of as a
particular example of warped Kaluza-Klein compactifications in string theory.
However, one of the main points of this paper is the claim
that there exists a one-parameter family of
spacetime varying vacua in IIB string theory, which naturally leads to an
exponential hierarchy and which at one point in the parameter space can be
identified with particular vacua of F-theory.

The singular properties
of the metric of our solution, responsible for the exponential hierarchy,
can be resolved in string theory via a non-supersymmetric
effect analogous to the \enh\ mechanism. This mechanism is based on a
repulsive nature of the naked singularity as seen by quantum probes of the
background. This last fact should be contrasted to the usual resolution
of classically repulsive singularities.
Our results offer an exciting possibility that supersymmetry breaking
is related to the emergence of an exponentially large hierarchy.

The paper is organized as follows:
       In section~2, we recall the construction of spacetime
varying string vacua and generalize
the Ans\"atze of Refs.~\cite{vafa, gh, FTh} for the
metric; we find non-trivial non-supersymmetric
solutions in which the moduli are
non-holomorphic functions over the noncompact space.
We concentrate on a single modulus scenario, in which the
modulus is given by the dilaton-axion system of $D=10$
type IIB string theory, even though we can in general treat the dynamics of
moduli arising from generic non-supersymmetric compactifications of type I,
type IIA,
or heterotic string theory.

In particular, we present a general solution
for the single modulus scenario in type IIB string theory in which the
axion-dilaton system is only
a function of the angular coordinate of the transverse
space. For simplicity we take the longitudinal part of
the metric for our solution to be conformally flat, even though a more
general class of solutions can be found.
We give a detailed account of the spacetime properties of this particular
kind of solution; the
source-free case corresponds to a neutral black 7-brane (analogous
to the Schwarzschild black hole in $D=4$),
while the charged brane,
which represents a
non-trivial axion/dilaton configuration, is obtained by adding
7-branes to the neutral background.

Our solution possesses a classical naked singularity
which is responsible for the existence of an exponential hierarchy.
We conduct both classical and quantum analysis of the effective dynamics of
a test probe in the background of such a naked singularity.
While the classical analysis indicates that the naked singularity
is attractive, the quantum treatment of the same problem reveals
that the singularity is reflective, and that the whole spacetime
looks like a box with impenetrable walls placed at the positions
of the classically naked singularities.

Finally
we study a supersymmetric limit of our solution and
its interpretation as a particular vacuum of F-theory.
We are guided by
the similarity between the transverse
parts of the corresponding metrics
for the supersymmetric and non-supersymmetric
solutions. We also point out
the crucial role played by the $SL(2,\ZZ)$ symmetry
of IIB string theory in establishing this limit.

In order to understand the supersymmetric limit of our solution as
well as the nature of the naked singularities from the string theory
point of view,
in section~3, we probe the geometry of our solution with $D7$-branes
while in section~4, we consider a non-supersymmetric effect
analogous to the \enh\ mechanism.
While physics is in general dependent on the nature
of the probe, we find that the results of the probe analysis are
in accord with the results of section~2: a quantum probe of our
background must remain in a scattered state although the effective
potential exerted by the classically naked singularity is
attractive. We repeat the probe analysis after wrapping our
ten-dimensional background on a $K3$, which is potentially
relevant from a purely phenomenological view-point, and
find that
the naked singularites while present in supergravity are resolved
in string theory in such a way that the exponential hierarchy
is preserved.

\section{The Cosmic Brane Solution}
In this section we review the solution of~\cite{bhmone} and
discuss in detail its most general form.

\subsection{The general problem}
Let us consider compactifications of string theory in which the
``internal'' space (a Calabi-Yau $n$-fold $M_n$)
varies over the ``observable'' spacetime. The parameters of the
``internal'' space then become spacetime variable moduli fields
$\f^\a$. The effective action describing the coupling of
moduli to gravity of the observable spacetime
can be derived by dimensionally reducing the higher dimensional
Einstein-Hilbert action~\cite{vafa,gh}. In this procedure one retains
the dependence of the Ricci scalar in the gravitational action only
on the moduli $\f^\a$. Thus, the relevant part of the low-energy
effective $D$-dimensional action of the moduli of the Calabi-Yau
$n$-fold, $M_n$, coupled to gravity reads~\footnote{In general the
action contains a scalar potential,
whose leading constant term represents an effective cosmological constant.
Thus, in principle it is possible to discuss solutions which interpolate
between Randall-Sundrum type of models and the solutions discussed below.}
\begin{equation}
       S_{\rm eff} = {1\over2\kappa^2}\int\rd^D x \sqrt{-g} ( R
               - G_{\a \bar{\beta}}g^{\mu \nu}
                 \vd_{\mu} \f^\a \vd_{\nu} \f^{\bar{\beta}}
                 +...)~,
\Label{e:effaction}
\end{equation}
where $\mu,\nu=0,{\cdots},D-1$, and $2\kappa^2=16\p G^D_N$, where $G_N^D$
is the $D$-dimensional Newton constant. We neglect higher derivative terms
and set the other fields in the theory to zero as in~\cite{vafa}.

We will restrict the moduli to depend on $x_i$,
$i{=}D{-}2,D{-}1$, so that $\vd_a \f{=}0$,
$a{=}0,{\cdots},D{-}3$.  The equations of motion are
\begin{equation}
       g^{ij} \Big(\nabla_{i} \nabla_{j} \f^\a
       + \Gamma^\a_{\beta \gamma} (\f, \bar{\f}) \vd _{i}
       \f^{\beta} \vd_{j} \f^{\gamma} \Big) = 0~,\Label{e:scalar}
\end{equation}
and
\begin{equation}
       R_{\mu\nu} - \inv2 g_{\mu\nu} R = - T_{\mu\nu} (\f,\bar{\f})~,
\Label{e:einstein}
\end{equation}
where the energy-momentum tensor of the moduli is
\begin{equation}
       T_{\mu \nu}
       = - G_{\a \bar{\beta}} \Big(\vd_{\mu} \f^\a\vd_{\nu}
        \f^{\bar{\beta}} - \inv2 g_{\mu \nu}\, g^{\rho\sigma}
        \vd_{\rho} \f^\a \vd_{\sigma} \f^{\bar{\beta}}\Big)~.
\end{equation}

It is useful to define $z \equiv (x_{D-2} + i x_{D-1})$ and
rewrite the effective action  as
\begin{equation}
       S_{\rm eff} = {1\over2\kappa^2}\int\rd^D x~ \sqrt{-g}\> R
                     - {1\over2\kappa^2}\int\rd^{D-2}x~ E~,
\end{equation}
where $d^{D-2}x$ refers to the integration measure over the first
$D-2$ coordinates, $x_0,{\cdots},x_{D-3}$. $E$, which we later interpret
as the energy density
(tension) of the cosmic brane, is given by
\begin{equation}
       E \equiv \int\rd^2z \sqrt{-g}~ G_{\a \bar{\beta}} g^{z \bar{z}}
       \Big(\vd_{z} \f^\a \vd_{\bar{z}} \f^{\bar{\beta}}
        + \vd_{\bar{z}}\f^\a\vd_{{z}}
        \f^{\bar{\beta}}\Big)~.\Label{e:Etension}
\end{equation}

We are interested in codimension-2 compactifications. Therefore we start
with the following Ansatz for the metric (writing $r\define|z|$)
\begin{equation}
       \rd s^2 = e^{2 A(r)} \eta_{ab} \rd x^a \rd x^b
       + e^{2B(r)} \rd z \rd\bar{z}~.\Label{e:metric}
\end{equation}
This type of Ansatz has appeared recently in various field
theory models~\cite{cohen,cp}, supergravity inspired
scenarios~\cite{CHT} and in the context of string theory~\cite{ein,dudas}.
The authors of~\cite{CHT} considered the possibility of having a
superpotential and hence a potential for the scalar fields. The
scalars in our effective action are assumed to be true moduli, with
no (super)potential.

In the remainder of this paper we will consider a single
modulus scenario, $\f^\a{=}\tau$ in which $\tau=a + ie^{-\F}$ is the
axion-dilaton system of the $D{=}10$ type~IIB string theory or
compactifications to lower dimensions. The holomorphic solution
$\tau=\tau(z)$ is that of
$D7$-branes~\cite{FTh}.

\subsection{The spacetime metric}
\Label{s:SpTmetric}
The general solution can be obtained
along the lines of our earlier analysis~\cite{bhmone}.
Note that the diffeomorphism invariance can be used to reduce the number
of integration constants in solving for the warp factors
$A$ and $B$:~\footnote{We
thank B.~Kol for illuminating discussions on this point.}
\begin{eqnarray}
      e^{2A(r)} &=& [1+a_0 \log(r)]^{{2\over D-2}}~,
      \Label{e:A}\\[1mm]
      e^{2B(r)} &=&
      l^2\,[1+a_0 \log(r)]^{-{D-3\over D-2}}\,r^{-2-\xi[a_0 \log(r)+2]}~.
      \Label{e:Bone}
\end{eqnarray}
We note that, in fact, both warp factors can be expressed in terms of the
harmonic function $Z_7(r)$:
\begin{eqnarray}
      e^{2A(r)} &=& [Z_7(r)]^{{2\over D-2}}~,\qquad
      [Z_7(r)]~\define~[1+a_0 \log(r)]~,
      \Label{e:AZ7}\\[2mm]
      e^{2B(r)} &=&
      l^2\,e^{\xi a_0^{-1}(1+1/\xi)^2}
      [Z_7(r)]^{-{D-3\over D-2}}\,e^{-\xi a_0^{-1}[Z_7(r)+1/\xi]^2}~,
      \Label{e:BZ7}
\end{eqnarray}
where the $\xi$-dependent factor appears akin to a screening
factor, such as in the screened Coulomb potential.
    Here the constant $l$ sets the length scale in the $z$-plane and is
required on dimensional grounds, and $\xi$ and $a_0$ are two integration
constants obtained when solving the Einstein equations~(\ref{e:einstein})
with a non-zero stress tensor
$T_{\mu\nu}$. In particular, $\xi$ determines the Ricci tensor
\begin{equation}
     R_{\mu\nu}=\hbox{\rm diag}
       [\,\underbrace{0,\cdots,0}_{D-1},2a_0\,\xi\,l^{-2}\,]
     \quad\hbox{for}\quad x^\mu=(t,x^1,\cdots,x^{D-3},r,l\q)~,
     \Label{e:Ricci}
\end{equation}
where the polar angle has been rescaled as $l\q$  for dimensional
reasons. Through
Einstein's equation, $\xi$ then also determines the strength of the stress
tensor:
\begin{eqnarray}
     [\,T_{\mu\nu}\,]&=&[\,T(r)\,\eta_{ab}\,]
          \oplus\hbox{\rm diag}[\,-a_0\xi r^{-2},a_0\xi\,]~, \\
     T(r) &=& -a_0\,\xi\, l^{-2}\,[1+a_0\log(r)]\,r^{\xi[2+a_0\log(r)]}~.
\end{eqnarray}

Furthermore, we note that $\xi$ and $a_0$ are not entirely independent (see
Eq.~(\ref{e:wReal}) below):
\begin{equation}
     \hbox{sign}(\xi)~=~\hbox{sign}(a_0)~, \quad\hbox{\ie}\quad
     \xi\,a_0\geq0~. \Label{e:signa0xi}
\end{equation}
In fact, since the Ricci 2-form, $R_{\mu\nu}\rd x^\mu\rd x^\nu$, is a
representative for the first Chern class, $c_1$, the
result~(\ref{e:Ricci}) together with~(\ref{e:signa0xi}) implies that
$c_1>0$ when $\xi\,a_0>0$. As is well known, this obstructs supersymmetry.
Now, although each of the three limits $a_0\to0$, $\xi\to0$ and
$l\to\infty$ removes this obstruction, we will
see in section~\ref{s:SuSy} that only the $a_0\to0$ limit leads to
supersymmetric solutions.

For future reference, we note that the metric~(\ref{e:metric}) with the
warp factors~(\ref{e:A}) and~(\ref{e:Bone}) is almost invariant under the
$r\to1/r$ transformation. To be precise, we change variables $r\to1/r$:
\begin{eqnarray}
     e^{2A(1/r)} &=& [1-a_0\log(r)]^{2\over D-2}, \nn\\[1mm]
     e^{2B(1/r)} &=& l^2\,[1-a_0\log(r)]^{-D-3\over D-2}\,
               r^{-2+\xi[2-a_0\log(r)]}~,
     \Label{e:ABInvr}
\end{eqnarray}
where the $r^{-2}$ factor in $e^{2B}$ does not change, owing to the way it
appears in the line element:
\begin{equation}
     \rd s^2~=~\ldots~+(\cdots
               r^{-2})(\rd r^2+r^2\rd\q^2)~,
\end{equation}
and that $r^{-1}\rd r$ and $\rd\q$ are invariant under the $r\to1/r$
transformation. Thus, we have that the inversion transformation
\begin{equation}
     {\cal I}\colon (r,a_0,\xi) ~\mapsto~ (1/r,-a_0,-\xi)
     \Label{e:InvSymm}
\end{equation}
is a symmetry of the metric~(\ref{e:metric}).

Finally, we note that the appearance of dimensionally unnormalized $r$ as
the argument of the logarithm in~(\ref{e:A}) and~(\ref{e:Bone}) indicates
that we have set the radial length unit to 1.

\bigskip\noindent$\underline{\hbox{Special properties of codimension-2}}$
\nobreak\vglue-1mm\noindent
Before we proceed, a few remarks are in order regarding the special
properties of codimension-2 solutions, \ie\ when the transversal space is
2-dimensional.

First, note the logarithmic dependence of the warp factors~(\ref{e:A})
and~(\ref{e:Bone}). This, of course, is the expected behavior of Coulomb
potentials, and (more generally) Green's functions of the Laplacian in
2-dimensional space.
Recall that in higher dimensions the corresponding quantities
diverge at one end and vanish at the other end of the natural domain of
the appropriately chosen radial variable, $r\in[0,\infty)$. This defines
the location of the `source,' at $r=0$, and the `asymptotic region' at
$r=\infty$. Unlike that, in 2 dimensions, the Green's functions diverge at
both ends, $r=0$ and $r=\infty$, implying that the `source' is in this
sense distributed at both of these locations. Note also that if the
Coulomb potential (Green's function) is attractive (goes to negative
infinity) at one end, it then is repulsive (goes to positive infinity) at
the other end.

This dimensionally atypical behavior also implies the absence of an obvious
candidate `asymptotic region' at either end. Instead, we note that the
logarithmic Coulomb potential necessarily vanishes at an intermediate
radius. In our case~(\ref{e:A}) and~(\ref{e:Bone}), this is just $\log(r)$
and it vanishes at $r=1$. For want of a better choice, we then treat the
region around $r=1$ as a surrogate for `the asymptotic region.' Note
however that, unlike in higher dimensions, the `force' determined by a
Coulomb potential in 2 dimensions does not vanish in the `surrogate
asymptotic region,' $r\sim1$.

In addition, the harmonic function $Z_7(r)$, as defined in Eq.~(\ref{e:A}),
exhibits another special point, $\rs=e^{-1/a_0}$, where it vanishes. This
is the location of a naked singularity; see in particular
sections~\ref{s:SpTimeProps} and~\ref{s:TrSpace}.

\bigskip
Having found a solution to the Einstein equations in the
form~(\ref{e:metric}), we now turn to a detailed discussion of the general
solution for $\tau$.

\subsection{The general solution for the toral modulus}
\Label{s:GenModulus}
The equation of motion~(\ref{e:scalar}) for the toral modulus, $\t=\t(\q)$,
becomes:
\begin{equation}
     \t'' + {2\over \bar\t-\t}(\t')^2 ~=~0~,
     \Label{e:GMEoM}
\end{equation}
where the coefficient of $(\t')^2$ is the Cristoffel connection,
$\G_{\t\t}^\t$, derived from the Teichm\"{u}ller metric; primes indicate
derivatives with respect to the argument.

By writing $\t=\t_1+i\t_2$ and separating the real and the imaginary parts
of Eq.~(\ref{e:GMEoM}), we obtain the following system of two coupled,
non-linear, second order ordinary differential equations:
\begin{eqnarray}
      \t_2\t_1'' &=& 2\t_1'\t_2'~, \Label{e:GM1}\\
      (\t_1')^2  &=& (\t_2')^2 - \t_2\t_2''~. \Label{e:GM2}
\end{eqnarray}
   From Eq.~(\ref{e:GM2}), it is immediate that
\begin{equation}
      \t_1 ~=~ b_0 \pm \int\rd\q\sqrt{(\t_2')^2 - \t_2\t_2''}~.
      \Label{e:tau1int}
\end{equation}
Therefore, given a solution for $\t_2(\q)$, Eq.~(\ref{e:tau1int}) gives the
corresponding $\t_1(\q)$.

\subsubsection{Type-I solution}
\Label{ss:1stSolution}
Here, we find the first branching point in the solution process. If the
left-hand side of Eq.~(\ref{e:GM2}) and so also the integrand in
Eq.~(\ref{e:tau1int}) vanishes, the solution may be written in the form
\begin{equation}
      \t_I(\q) ~=~ b_0 + i\,g_s^{-1}\,e^{\w(\q-\q_0)}~,
      \qquad b_0,g_s,\w,\q_0 = \const~,
      \Label{e:tauI}
\end{equation}
and where $g_s$ is the string coupling constant.
Now, the Einstein equation~(\ref{e:einstein}) provides a relation between
the constants in $\t_I$ and those in the metric~(\ref{e:metric}),
with~(\ref{e:A}) and~(\ref{e:Bone}). In particular,
\begin{equation}
    \w=\sqrt{8\xi a_0}~,\qquad\hbox{\it i.e.,}\qquad \w^2=8\xi a_0~\geq0~,
    \Label{e:wReal}
\end{equation}
where the last inequality is enforced by the fact that $\w$ must be real.
Also, $e^{\w \theta_0}$ can be reabsorbed in
$g_s^{-1}$; equivalently, we set $\q_0=0$ and fix the defining domain of
$\q$ to be $[-\p,+\p]$. Note furthermore that the constants $b_0$ and
$g_s$ are chosen such that rotations around the origin, \ie\ varying $\q$
through its defining domain induces an $SL(2,\ZZ)$ action on this
solution~\footnote{By imposing a covariance under $SL(2,\ZZ)$, rather
than $SL(2,\IR)$, we expect our solution to be a string vacuum
rather than just a (super)gravity solution.}:
\begin{equation}
        \t_I(+\p)=\frac{a\,\t_I(-\p)+b}{-\t_I(-\p)+d}~.
        \Label{e:tau1tr}
\end{equation}
Thus, in general only very special values of the string coupling will
be allowed. We will return to this point in section~\ref{s:SuSy}
  when we discuss
the supersymmetric and F-theory limit of our solutions.

\subsubsection{Type-II solution}
\Label{ss:2ndSolution}
On the other hand, if the left-hand side of Eq.~(\ref{e:GM2}) does not
vanish, this equation can be used to express $\t_1(\q)$ in terms of
$\t_2(\q)$, as done in Eq.~(\ref{e:tau1int}). In this case we have:
\begin{equation}
    \t_{II}(\q)=b_0 \pm
                   e^{\a_0}{\sinh[\w(\q-\q_0)]+i\over\cosh[\w(\q-\q_0)]}~,
    \qquad b_0,e^{\a_0},\w,\q_0=\const
\end{equation}
We may again fix $\q_0=0$ and choose the defining domain $\q\in[-\p,+\p]$.
As for $\t_I$, the Einstein equation~(\ref{e:einstein}) enforces the
relation
$\w=\sqrt{8\xi a_0}$. Finally, since
\begin{equation}
     \Imm(\t_{II}) = \pm e^{\a_0}{i\over\cosh[\w(\q-\q_0)]} = e^{-\F}~,
\end{equation}
we are forced to identify $e^{\a_0}=g_s^{-1}$, so
\begin{equation}
    \t_{II}(\q)=b_0 \pm
                   g_s^{-1}\,{\sinh[\w(\q-\q_0)]+i\over\cosh[\w(\q-\q_0)]}~,
    \qquad b_0,g_s,\w,\q_0=\const
    \Label{e:tauII}
\end{equation}

Having exhausted all the logical possibilities in the solution process, we
conclude that the two branches of solutions, $\t_I(\q)$ and $\t_{II}(\q)$,
as given in Eqs.~(\ref{e:tauI}) and~(\ref{e:tauII}) respectively,
represent the most general solutions to the modular equation of
motion~(\ref{e:GMEoM}).

However, there is also the trivial solution, $\t=\const$. Apart from
satisfying~(\ref{e:GM1}--\ref{e:GM2}), this solution can also be obtained
as a special limit either of $\t_I$ or $\t_{II}$. 
When $g_s^{-1}\to 0$ both
$\t_I$ and $\t_{II}$ become equal to $b_0$, which follows from
Eqs.~(\ref{e:tauI}) and (\ref{e:tauII}). In addition, by letting $\xi\to
0$, Eq.~(\ref{e:wReal}) implies that $\t_I,\,\t_{II}\to b_0$.

Finally, we note that it is possible to choose $b_0$ and $e^{\a_0}$ such
that rotations around the origin, \ie\ varying $\q$ through its defining
domain induces a monodromy action on
$\t_{II}(\q)$:
\begin{equation}
        \tau_{II}(+\pi)=\tau_{II}(-\pi)\pm n~,
        \Label{e:tau2tr}
\end{equation}
where $n$ is an integer. We will return to the interpretation of $n$ when
discussing the supersymmetric limit in~section~\ref{s:SuSy2}.

It is remarkable that although both solutions, $\t_I(\q)$ and
$\t_{II}(\q)$, are strictly aperiodic functions of $\q$, it is nevertheless
possible to ensure that they are $SL(2,\ZZ)$-covariant, as given in
Eqs.~(\ref{e:tau1tr}) and~(\ref{e:tau2tr}), respectively.

\subsection{Spacetime properties}
\Label{s:SpTimeProps}
In the model considered above, where the dilaton and axion vary over
$z{=}x_8{+}ix_9$, the metric~(\ref{e:metric}), with warp
factors~(\ref{e:A}) and~(\ref{e:Bone}) and for $\xi=-1$, may be written as
\begin{equation}
    \rd s^2~=~\Big({\rho\over\rho_0}\Big)^{2\over D-2}\eta_{ab}\rd x^a\rd x^b
       + l^2 \Big({\rho_0\over\rho}\Big)^{D-3\over D-2}
       e^{({\rho_0}^2 -  \rho^2)/{\rho_0}} \Big(\rd\rho^2 + \rd\q^2\Big)~,
      \Label{e:metricrho11}
\end{equation}
which is identical to that of the global cosmic brane
solution studied by Cohen and Kaplan in~\cite{cohen}; we have used
$\rho \define |a_0|^{-1}[1-|a_0| \log(r)]$ and $\rho_0\define
|a_0|^{-1}=\rho|_{r=1}$, and $l$ is the length scale introduced in
Eq.~(\ref{e:Bone}) for dimensional reasons.

In the remainder of this section
we would like to study the spacetime properties
of the general solution~(\ref{e:A}--\ref{e:Bone}) in more detail.
In particular, we want to answer
questions about the stability of the solution and the relation to its
supersymmetric cousins. To this end, we first look at the special case
when $\xi=0$.

\subsubsection{The source-less case}
The relation $\w=\sqrt{8\xi a_0}$ then implies that also
$\w=0$, so both $\t_I(\q)$ and $\t_{II}(\q)$ become constant, and
consequently the stress tensor, $T_{\mu\nu}$, now vanishes.
    The source-less Einstein equations
\begin{equation}
R_{\mu\nu} - {1\over 2} g_{\mu\nu} R = 0
\Label{e:einsteinzero}
\end{equation}
are solved by the Ansatz~(\ref{e:metric}), with the warp
factors~(\ref{e:A}) and~(\ref{e:Bone}) unchanged, except for setting
$\xi=0$:
\begin{equation}
e^{2A(r)} = [1+a_0\log(r)]^{2\over D-2},\,\quad
e^{2B(r)} = l^2\,[1+a_0\log(r)]^{-{D-3\over D-2}}\,r^{-2}~.
\Label{e:generalwarp}
\end{equation}

It is easy to check that $R=0$ and hence $R_{\mu\nu}=0$. Yet, the warp
factors $e^{2A(r)}$ and $e^{2B(r)}$, respectively, vanish and diverge at
$\rs=e^{-1/a_0}$.
    In that sense, this solution is similar to the Schwarzschild
solution~\footnote{For a recent review see~\cite{Peet}.}.
Indeed, the quadratic curvature scalar
\begin{equation}
     Riem^2 \define R_{\mu\nu\rho\sigma} R^{\mu\nu\rho\sigma}
\sim a_0^4\,l^{-4}\,[1 + a_0\log(r)]^{-2{D-1\over D-2}}~,
\end{equation}
diverges at $\rs=e^{-1/a_0}$. However, as opposed to the case of black
holes, here the horizon\footnote{Throughout this article, we will refer
to what is more precisely called the `putative horizon,' defined as the
place where the lapse function ($g_{tt}$, for a diagonal metric)
vanishes; see e.g.\ \S~2.3 of Ref.~\cite{Visser}.} coincides with this
singularity, which then is a naked (null) singularity.

In addition, at $r=0$ for $a_0<0$ ($r=\infty$ for $a_0>0$) the Killing
fields ${\rd\over\rd x^a}$ along the brane diverge, \ie\ there is a
time-like singularity at the origin (infinity)~\footnote{We thank D.~Marolf
for a discussion of this point.}.
Furthermore, in the limit when $a_0\to 0$ we recover
a flat space solution though with nontrivial topology, $R^{1,8}\times S^1$.
Therefore, it seems tempting to continue the analogy with the
Schwarzschild solution, in which the physical singularity vanishes if the
mass $M$ of the black hole is taken to zero.

In order to explore this analogy further, consider the
weak-field linearized Einstein equation~\cite{Visser,Weinberg}:
\begin{equation}
     \Big(R_{\mu\nu}-{1\over2}g_{\mu\nu}\,R\Big)_{\rm weak-field}~\approx~
     \Box\Big(h_{\mu\nu}-{1\over2}\eta_{\mu\nu}\,h\Big)
     ~\define~ \Q_{\mu\nu},   \Label{e:weakfield}
\end{equation}
where
\begin{equation}
     h_{\mu\nu}~=~g_{\mu\nu}-\eta_{\mu\nu}~,\qquad\hbox{and}\qquad
     h~=~\eta^{\mu\nu}h_{\mu\nu}~.
\end{equation}
Using the general form of our metric~(\ref{e:metric}), we find that
\begin{eqnarray}
     h_{\mu\nu}&=&\hbox{diag}\Big[(1{-}e^{2A}),
      \underbrace{(e^{2A}{-}1),{\cdots},(e^{2A}{-}1)}_{D-3},
      (e^{2B}{-}1),(e^{2B}{-}1)\Big]~,\quad\hbox{and} \nn\\
     h&=&(D{-}2)(e^{2A}{-}1)+2(e^{2B}{-}1)~, \qquad
        \hbox{for}\quad x^\mu=(t,x^1,{\cdots},x^{D-1})~.
\end{eqnarray}
Since $A$ and $B$ depend only on $r$, we substitute
$\Box\to {1\over r}{\rd\over\rd r}r{\rd\over\rd r}$. This and the
functional form of the warp factors~(\ref{e:A}) and~(\ref{e:Bone})
suggests using $\rho=a_0^{-1}[1{+}a_0\log(r)]$ as an abbreviation;
then ${\rd\over\rd\rho}=r{\rd\over\rd r}$.
    The definition~(\ref{e:weakfield}) then produces
\begin{eqnarray}
     \Q_{tt}&=&{1\over r^2}\Big({\rd^2\over\rd\rho^2}e^{2B}
                    + {D{-}4\over2}{\rd^2\over\rd\rho^2}e^{2A}
                    \Big)~, \Label{e:PsTgen}\\[2mm]
    &=&{4l^2\over r^4}[1{+}a_0\log(r)]^{-{D-3\over D-2}}\,C(r)
      ~-~\Big({D{-}4\over D{-}2}\Big)^2\,{a^2_0\over r^2}
      [1{+}a_0\log(r)]^{-2{D-3\over D-2}}~,\Label{e:PsTBscreen}\\[2mm]
    C(r)&=&1+\Big({D{-}3\over D{-}2}\Big)\,a_0[1{+}a_0\log(r)]^{-1}
            +{(D{-}3)(2D{-}5)\over4(D{-}2)^2}\>a_0^2[1{+}a_0\log(r)]^{-2}~.
     \Label{e:PsTnosrc}
\end{eqnarray}
The first term in~(\ref{e:PsTBscreen}) clearly diverges worse than the
second, and for two different reasons. For one thing, the exponent of the
logarithmic part of $e^{2B}$ is less than that of $e^{2A}$.
What is more important, however, $e^{2B}$ contains a `screening' factor,
$r^{-2}$, which is the source of the $r^{-4}$ factor in
the first term of~(\ref{e:PsTBscreen}).

Near $r=1$, where the weak-field approximation ought to apply, we
may neglect the logarithms, which leaves
\begin{equation}
     \Q_{tt}\sim{4l^2\over r^4}\,
     \Big[1+\Big({D{-}3\over D{-}2}\Big)\,a_0
            +{(D{-}3)(2D{-}5)\over4(D{-}2)^2}\>a_0^2\Big]
      ~-~\Big({D{-}4\over D{-}2}\Big)^2\,{a^2_0\over r^2}~.
\end{equation}
In comparison, the energy-momentum ``pseudo-tensor'' for the
Schwarzschild black hole solution is~\cite{Weinberg}
\begin{equation}
     \Q^{BH}_{tt} \sim {M\over r^3}.
\end{equation}

Thus, just as the black hole disappears when $M\to 0$, one might naively
expect the black 7-brane solution to vanish when $a_0\to 0$. This is in
fact not quite the case, as Eqs.~(\ref{e:PsTBscreen})--(\ref{e:PsTnosrc})
imply:
\begin{equation}
     \lim_{a_0\to0} \Q_{tt} ~\to~{4l^2\over r^4}~. \Label{e:BackGrnd}
\end{equation}
This term can be traced to the `background' source of $r^{-2}$ in
$e^{2B}$, indicating a (perhaps puzzling) non-trivial source. We wish to
emphasize that this `background' source is not unambiguously located at
$r=0$ only: the change of variables $r\to1/r$ leaves $e^{2B}\rd r^2$
invariant, except for the change $a_0\to-a_0$. So, in a sense, this
`background' source is located both at $r=0$ and at $r=\infty$.

To explore this further, we now include explicit sources in the analysis.

\subsubsection{The case with sources}
Let us consider a non-zero $T_{\mu\nu}$.  The warp factors of the
metric~(\ref{e:A}) and~(\ref{e:Bone}) read
\begin{eqnarray}
     e^{2A(r)}&=&[1+a_0\log(r)]^{2\over D-2}, \nn\\[1mm]
     e^{2B(r)}&=&l^2\,[1+a_0\log(r)]^{-D-3\over D-2}\,
               r^{-2-\xi[2+a_0\log(r)]}~,
     \Label{e:ABa0pos}
\end{eqnarray}
where we fix, for the time being, $a_0\geq0$.
    By taking the limit $\xi\to 0$, we recover the
solution~(\ref{e:generalwarp}) to the source-free Einstein equations.
Furthermore we compute the scalar curvature:
\begin{equation}
     R\sim \xi\,a_0\,l^{-2}\,r^{\xi[2+a_0\log(r)]}\,
           [1+a_0\log(r)]^{{D-3\over D-2}}~,
\Label{e:scurv.src}
\end{equation}
the square of the Ricci tensor:
\begin{equation}
     Ric^2 \sim \xi^2\,a_0^2\,\,l^{-4}
             r^{2\xi[2+a_0\log(r)]}\,[1+a_0\log(r)]^{2{D-3\over D-2}}~,
\Label{e:ricci2.src}
\end{equation}
and the square of the Riemann tensor:
\begin{eqnarray}
     Riem^2 &\sim& {\cal R}(r)\,a_0^4\,l^{-4}\,
             r^{2\xi[2+a_0\log(r)]}\,[1+a_0\log(r)]^{-2{D-1\over D-2}}~.\nn\\
     {\cal R}(r)&=&\textstyle ({(D-1)(D-3)\over(D-2)^2}
              -4{(D-3)\over(D-2)}\,\xi\,a_0\,\rho^2
              +4{D\over(D-2)}\,\xi^2\,a_0^2\,\rho^4)~,
\Label{e:riem2.src}
\end{eqnarray}
where we again used the abbreviation $\rho\define a_0^{-1}[1+a_0\log(r)]$.

Note that $Ric^2$ and $R$ provide no new information: $Ric^2$ behaves just
like the $\xi^2$ term in $Riem^2$, and the scalar curvature simply
behaves like $\sqrt{Ric^2}$.
    Just as for the source-free solution, $Riem^2$ diverges at
$\rs=e^{-1/a_0}$. (Although non-zero in general, $R$ and $Ric^2$
actually do vanish at $\rs$.)

In complete analogy with the source-less case, we compute the weak-field
linearization of the Einstein tensor in this situation
\begin{eqnarray}
     \Q_{tt}^\xi
    &=&{4l^2\over r^4}[1{+}a_0\log(r)]^{-{D-3\over D-2}}\,
          r^{-\xi[2{+}a_0\log(r)]}\,\cC_\xi(r)
      ~-~\Big({D{-}4\over D{-}2}\Big)^2\,{a^2_0\over r^2}
      [1{+}a_0\log(r)]^{-2{D-3\over D-2}}~,\nn\\[2mm]
    \cC_\xi(r)&=&\Big[\Big(1+\xi[1+a_0\log(r)]\Big)^2+
    {a_0\xi\over2}\Big]+\nn\\
    &&\quad\Big({D{-}3\over D{-}2}\Big)\,a_0[1{+}a_0\log(r)]^{-1}
     \Big(1+\xi[1+a_0\log(r)]\Big)+\nn\\
    &&\qquad{(D{-}3)(2D{-}5)\over4(D{-}2)^2}\>a_0^2[1{+}a_0\log(r)]^{-2}
      ~.\qquad
     \Label{e:PsTsrc}
\end{eqnarray}
Again, in the weak-field regime (near $r=1$), we neglect the logarithms,
which leaves
\begin{eqnarray}
     \Q_{tt}^\xi
    &\sim&{4l^2\over r^4}r^{-2\xi}\,\cC_\xi(1)
      ~-~\Big({D{-}4\over D{-}2}\Big)^2\,{a^2_0\over r^2}~,\nn\\[2mm]
    \cC_\xi(1)&=&\Big[(1+\xi)^2+{a_0\xi\over2}\Big]
               +\Big({D{-}3\over D{-}2}\Big)\,a_0(1+\xi)
               +{(D{-}3)(2D{-}5)\over4(D{-}2)^2}\>a_0^2~.
\end{eqnarray}
   From this, we furthermore find
\begin{equation}
     \lim_{a_0\to0+}\Q_{tt}^\xi\sim4l^2\,(1+\xi)^2\,r^{-2(1+\xi)-2}~,
     \Label{e:Qttlim}
\end{equation}
which is to be compared with the corresponding expression for the
supersymmetric D7-branes of Refs.~\cite{ein}. In fact,
the metric~(\ref{e:metric}) can be specialized so as to correspond to the
result of Refs.~\cite{ein}, by setting
\begin{equation}
     e^{2A}~\to~1~, \quad\hbox{and}\quad e^{2B}~\to~[1+\l\log(r)]\,r^{2B_1}~.
\end{equation}
In addition, $\t\to-\l+i[1{+}\l\log(r)]$. From these, we compute
\begin{equation}
     \Q_{tt}=-\l\,\delta(r)\,r^{2B_1} + 4B_1\l r^{2B_1-2}
              +4B_1^2[1+\l\log(r)]\,r^{2B_1-2}~.
\end{equation}
Thus,
\begin{equation}
     \lim_{\l\to0}\Q_{tt} \sim 4B_1^2\,r^{2B_1-2}~,
\end{equation}
which agrees with~(\ref{e:Qttlim}) upon the identification
$B_1={-}(\xi{+}1)$ when $a_0,\xi\geq0$.

The analogous result for the $a_0\leq0$ case is now easy to deduce, owing
to the inversion symmetry~(\ref{e:InvSymm}). We now obtain
\begin{equation}
     \lim_{a_0\to0-}\Q_{tt}^{-\xi}\sim4l^2\,(1-\xi)^2\,r^{-2(1-\xi)-2}~,
\end{equation}
and therefore $B_1={-}(1{+}|\xi|)$ in general.

On one hand, our $a_0$ and the $\l$ in Refs.~\cite{ein} appear in an
analogous, but not identical, fashion in the metric. In addition, the
functional dependence of both the axion and the dilaton in our
configurations is different from that in Refs.~\cite{ein}. However, our
solutions agree with that of Refs.~\cite{ein} only in the simultaneous
limit $a_0,\l\to0$. Also, comparison with Ref.~\cite{vafa} shows that
this indicates (in the present limit) the presence of $12(1+|\xi|)$
supersymmetric 7-branes. The extra `background' sources, remaining in the
$\xi\to0$ limit, have already shown up in our calculations for the
source-less case, in Eq.~(\ref{e:BackGrnd}). In this sense, the source-less
solution is akin to the Schwarzschild metric, where there is a
singularity without a corresponding `matter' ($T_{\mu\nu}$) source.

This analysis indicates that, of the limits discussed in
section~\ref{s:SpTmetric}, it is the $a_0\to0$ limit that makes our
solutions supersymmetric. We will discuss this in more detail in
section~\ref{s:SuSy}.

\subsubsection{The nature of the naked singularity}
While having a naked singularity is not desirable from a purely
    classical point
of view, it is possible that the singularity may be resolved in a
string theory context~\footnote{A
simple example of that is the supersymmetric
7-brane in supergravity which has a naked singularity at a finite distance
away from the origin. However, this UV-problem is resolved when considering
the $D7$-brane in string theory, because of the $SL(2,\ZZ)$ covariance.}.
Recently, examples of so called ``repulsons'',
a repulsive naked singularity~\cite{KL}, have been
studied in supergravity. These were found to be harmless once they
were considered in the dual gauge theory
formulation~\cite{joep,
alex,cliff,clifford,lerda} --- this is the \enh\ mechanism.
    It is therefore important to consider the classical
nature of the above naked singularity
as a first step towards answering the question whether there exists
a non-supersymmetric effect
analogous to the \enh\ mechanism in our case.

Let us
compute the time it takes a test particle to fall into, or get
reflected by the naked singularity. Following Kallosh and Linde~\cite{KL}
(see
also~\cite{cohen},\cite{behrndt} and~\cite{LL}) we have for a
particle with mass $m$, energy $E$, and angular momentum $L$,
\begin{equation}
     t~=~E\int\rd r {\sqrt{g_{rr}}\over g_{00}}\Big ({E^2\over g_{00}} -
g_{rr}^{-1}{L^2\over r^2} - m^2\Big )^{-1/2}~. \Label{e:time}
\end{equation}
As discussed in section~\ref{s:SpTmetric}, the only reasonable candidate
for an asymptotic region is around $r=1$. This is where the quantities
$E,L,m$ need to be evaluated for the equation~(\ref{e:time}).

To decide whether the singularity is repulsive or attractive (and
hence whether spacetime is geodesically complete or not) we have to see
if  the denominator of the integrand
has a zero. This would correspond to the test
particle coming to a stop and hence getting reflected from
the singularity. For
simplicity let us set $L=0$. The answer only depends on the behavior of
$g_{00}$ close to the singularity. Since $g_{00}\to 0$, the particle
does not turn around. From this equation, one sees that the
time the test particle
takes to reach the naked singularity is finite. Hence our
solution is geodesically incomplete and the effective potential is
attractive. Note that the Schwarzschild solution shares the same features
with our solution.

However, a quantum mechanical treatment of this problem shows that the
potential is so attractive that there are no normalizable bound states.
Hence, the particle must be reflected, \ie\ remain a superposition of
scattering states~\cite{cohen,bhmone}. The explicit treatment of a quantum
scalar particle probe of our background was done in Appendix~A
of~\cite{bhmone}, and will not be repeated here. Suffice it here to say
that the wave functions of the probe vanish as the singularities are
approached. Thus, effectively, the whole spacetime looks like a box with
impenetrable walls at the positions of the singularities. Finally, although
we used a scalar probe in Ref.~\cite{bhmone}, the analysis there suggests
that this qualitative feature is independent of the scalar nature of the
probe. Herein, we will assume this to be true.

\subsection{The transverse space}
\Label{s:TrSpace}
In order to understand the effects of the charged 7-branes in our
solution it is important to study their effect on the transverse,
$z$-space. We first note that the longitudinal part of
spacetime, spanned by the coordinates $t,x^1,{\cdots},x^{D-3}$, is fibered
over the transverse space. For any point on the base the fiber is just a
$(D{-}2)$-dimensional Minkowski space. However, because of the behavior
of the warp factor $e^{2A}$ in the metric~(\ref{e:metric}), the distances
within this $(D{-}2)$-dimensional Minkowski space vary over the transverse,
$z$-space. In particular, all distances and in fact the volume of the
entire $(D{-}2)$-dimensional Minkowski space  vanishes at the naked
singularity.
    When we wrap this solution on a compact space, such as
$K3$, we will see that although the supergravity solution breaks down
when $V_{K3}\to 0$, string theory is still well-defined.

\subsubsection{Physically permissible regions}
\Label{s:permit}
Owing to the appearance of the logarithm and the fractional powers in the
warp factors~(\ref{e:A}) and~(\ref{e:Bone}), not all values of
$r\in[0,+\infty)$ are physically admissible. In fact, depending on the
sign of $a_0$, there appear two physically permissible regions of the
transverse space.

First of all, the naked singularity occurs at $\rs^+\define
e^{-1/a_0}=e^{-1/|a_0|}$ for $a_0>0$, but at
$\rs^-\define e^{-1/a_0}=e^{+1/|a_0|}$ when $a_0<0$. It should be clear
that
\begin{equation}
     \rs^+<\rs^-\quad\hbox{for}\quad|a_0|<\infty~,\quad\hbox{and}\quad
     \lim_{a_0\to+\infty}\rs^+=1=\lim_{a_0\to-\infty}\rs^-~.
     \Label{e:overlap}
\end{equation}

For $a_0>0$, the warp factors, $e^{2A}$ and $e^{2B}$ and so also the
metric~(\ref{e:metric}) are all finite, real and positive as long as
$r\in[\rs^+,\infty)$.
    However, for $r<\rs^+$ the warp factors become complex:
$e^{2A}=|e^{2A}|e^{2i\p\over D-2}$ and
$e^{2B}=|e^{2B}|e^{-{D-3\over D-2}i\p}$.
    Therefore, when $a_0>0$ only $\rs^+\leq r<\infty$ is physical.
    We refer to this branch as the `outside' region of the transverse space.

    On the other hand, when $a_0<0$, the warp factors, $e^{2A}$ and
$e^{2B}$, and so also the metric~(\ref{e:metric}) are all finite, real and
positive while $r\in(0,\rs^-]$.
    However, the warp factors now become complex
($e^{2A}=|e^{2A}|e^{2i\p\over D-2}$ and
$e^{2B}=|e^{2B}|e^{-{D-3\over D-2}i\p}$) for $r>\rs^-$.
    Therefore, when $a_0<0$ only $0<r\leq \rs^-$ is physical.
This complementary branch is naturally identified as the `inside' region of
the transverse space.

Clearly, the inversion symmetry~(\ref{e:InvSymm})
maps the `outside' and the the `inside' region into one another.
Both of these regions have the topology of $\IR_+{\times}S^1$,
containing the boundary circle with $r=\rs^\pm$. (To be pedantic, the
`inside' region of the transverse space is $\IR_-{\times}S^1$, but
topologically $\IR_+\cong\IR_-$.) Owing to the
relations~(\ref{e:overlap}), the two regions overlap in the annulus
$\rs^+\leq r\leq \rs^-$. We may then consider a candidate transverse
space manifold obtained by gluing the two regions together into an `open'
annulus: the $z$-plane with the origin excised. This has the topology
of $\IR^1{\times}S^1$.

In this article, we focus on the two separate physically permissible
regions of the transverse space: `outside' with $a_0>0$ and
$r\in[\rs^+,\infty)$, and `inside' with $a_0<0$ and $r\in(0,\rs^-]$, and
their respective $a_0\to0$ limits.

\subsubsection{The source-less case}
Again, we start by examining the source-less case,
and focus on the transverse part of the metric.
In terms of the standard polar coordinates we have,
\begin{equation}
     \rd s_{\perp}^2 = l^2\,[1+a_0 \log(r)]^{-{(D-3)\over(D-2)}}\,
                       r^{-2}(\rd r^2 + r^2\rd\q^2)~.
\end{equation}
In writing the metric in this form we have
used the diffeomorphism invariance for the longitudinal part (which
removes all independent integration constants for $e^{2A}$
and fixes the overall scaling factor for $e^{2B}$).
Thus, we are left with one free parameter, $a_0$, which
can be either positive or negative.

The transverse metric can be rewritten in the standard form,
\begin{equation}
     \rd s_{\perp}^2 = \rd p^2 + P(p)^2 \rd\q^2~,
\end{equation}
so that
\begin{equation}
     \rd p~=~l\,[1+a_0 \log(r)]^{-{(D-3)\over2(D-2)}}\,r^{-1}\rd r~.
     \Label{e:dpnosrc}
\end{equation}
When integrating this equation, it remains to determine the lower
limit of the $r$-integration on the left hand side. Since the natural
boundary of $r$ is $e^{-1/a_0}$ (that is, $\rs^+$ in the `outside' region,
and $\rs^-$ in the `inside' region), it seems reasonable to use this as the
reference point when integrating the right hand side of~(\ref{e:dpnosrc}).
However, with the benefit of hindsight, we will leave this lower limit,
$r_0$, unspecified for now and introduce a corresponding constant. With
this in mind, we find the proper distance to be
\begin{equation}
     p~=~\tilde p_0 + {2l\over a_0}{D-2\over D-1}\,
         \Big([1+a_0\log(r)]^{D-1\over2(D-2)}-\tilde\l\Big)~,
     \Label{e:PD.nosrc}
\end{equation}
where the constants in the $p$- and $r$-integration are $\tilde p_0$ and
$\tilde\l$, respectively. In fact, since we restrict $r$ to the physically
permissible region, $[\rs^+,+\infty)$ when $a_0>0$ and
$(0,\rs^-]$ when $a_0<0$,
\begin{equation}
     \tilde\l~\define~[1+a_0\log(r_0)]^{D-1\over2(D-2)}
     \quad\To\quad \tilde\l\geq0~.   \Label{e:twlambda}
\end{equation}
The naked singularity is now located at
\begin{equation}
     p_{\rm s}\define p(\rs) ~=~ \tilde p_0 - {2l\tilde\l\over
a_0}{D-2\over D-1}~.
     \Label{e:psing}
\end{equation}
Depending on the sign of $a_0$, the qualitative features of $p(r)$ vary.
We discuss in turn these two different proper distances. In
particular, note that
\begin{equation}
     p_{\rm s}^+=\tilde p_0-{2l\tilde\l\over|a_0|}{D-2\over D-1} ~<~
     p_{\rm s}^-=\tilde p_0+{2l\tilde\l\over|a_0|}{D-2\over D-1}~.
\end{equation}
It should be clear that while $\tilde p_0$ shifts the positions of
$p_{\rm s}^+$ and $p_{\rm s}^-$ {\it together\/}, $\tilde\l$ controls
the splitting
between $p_{\rm s}^+$ and $p_{\rm s}^-$. As long as we discuss either
one of the two
regions of the transverse space, using two constants is redundant and we
may fix $\tilde p_0=0$; $\tilde\l$ then controls $p_{\rm }^\pm$, the position
of the naked singularity.

For $a_0>0$, $p=p^+(r)$ is real, and varies as
$p^+(r)\in[\tilde p_0,+\infty)$ when $r\in[\rs{=}e^{-1/a_0},\infty)$, \ie\
in the `outside' physical region.
    However, in the unphysical region $r<\rs$, the proper distance $p^+(r)$
becomes complex and of the form $|p^+|e^{i{D-1\over2(D-2)}\pi}$, just as do
the warp factors and the whole metric~(\ref{e:metric}). This then further
justifies our assessment that $0<r<\rs^+$ is unphysical when $a_0>0$.
    Note that this `outside' region of the transverse space is parametrized
by $p^+(r)\in[p_{\rm s}^+,+\infty)$, and so includes the boundary circle at
$p=p_{\rm s}^+$, where the naked singularity is located.

    On the other hand, when $a_0<0$, $p=p^-(r)$ is real, and varies as
$p^-(r)\in(-\infty,\tilde p_0]$ when
$r\in(0,\rs{=}e^{-1/a_0}]$, \ie\ in the `inside' physical region.
    However, for $r>\rs$ the proper distance function $p^-(r)$ becomes complex
of the form $|p^-|e^{i{D-1\over2(D-2)}\pi}$, just as do the warp
factors and the metric. This again justifies our earlier conclusion that
$\rs^-<r<\infty$ is unphysical when $a_0<0$.
    This `inside' region of the transverse space is parametrized by
$p^-(r)\in(-\infty,p_{\rm s}^-]$, and so includes the location of the naked
singularity: the boundary circle at $p=p_{\rm s}^-$.

Since $\tilde\l\geq0$, the two branches overlap for $p_{\rm
s}^+<p<p_{\rm s}^-$. Then,
for suitably chosen $\tilde\l$, the two functions $p^\pm(r)$ will have one
or two common points (see Fig.~\ref{f:p(r)}). In fact, for
$\tilde\l=1$, the two proper distances not only meet at $r=1$ (with
$p^+(1)=\tilde p_0=p^-(1)$), but are in fact tangential to each other at
that point.
\begin{figure}[ht]
       \framebox{\epsfxsize=160mm%
       \hfill~% Here be picture.
       \epsfbox{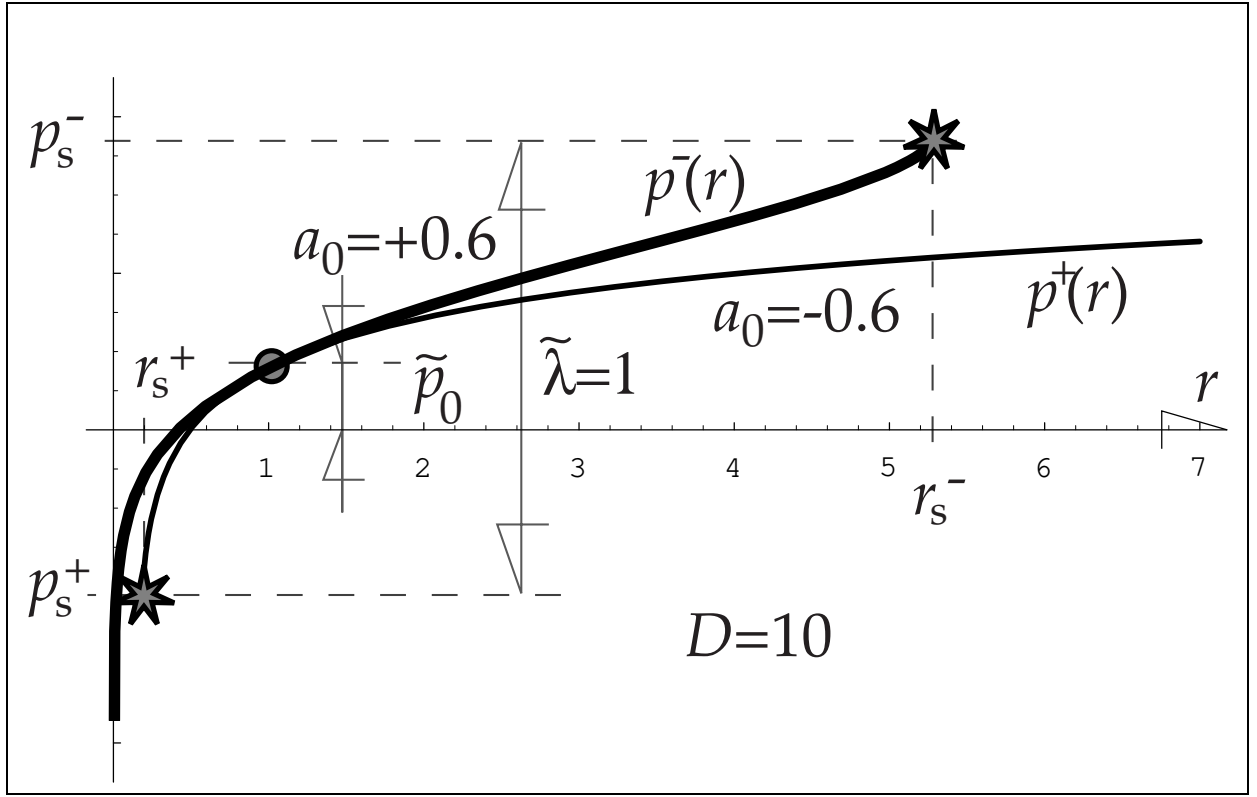}
       \hfill~}
       \caption{A plot of the two branches of the proper distance, $p^+(r)$
       and $p^-(r)$, as functions of the (flat) auxiliary distance, $r$.}
       \Label{f:p(r)}
\end{figure}
    It is then possible to construct a manifold by
gluing the two patches together along the `unit' circle ($r=1$ and
$p=\tilde p_0$). This new transverse space would then be parametrized by
$p^-\in(-\infty,\tilde p_0]$ and $p^+\in[\tilde p_0,+\infty)$.
Topologically, this is a cylinder, $\IR^1{\times}S^1$. Owing to the
smooth joining of $p^+$ and $p^-$ at $r=1$, the mapping from the
$(r,\q)$-plane (minus the origin) to the $(p,\q)$-cylinder is not only
continuous but also smooth.

Next, we compute
\begin{equation}
     P(p)~=~l\,\Big(\tilde\l+{a_0\over2l}{D-1\over D-2}(p-\tilde p_0)\Big)
                ^{-{D-3\over D-1}}.
     \Label{e:CF.nosrc}
\end{equation}
Note that at the naked singularity,
$p_{\rm s}=\tilde p_0-{2l\tilde\l\over a_0}{D-2\over D-1}$ and
$P(p_{\rm s})=\infty$; see Fig.~(\ref{f:P(p)}). On the other end,
$P(\pm\infty)=0$.
\begin{figure}[ht]
       \framebox{\epsfxsize=160mm%
       \hfill~%Here be picture.
       \epsfbox{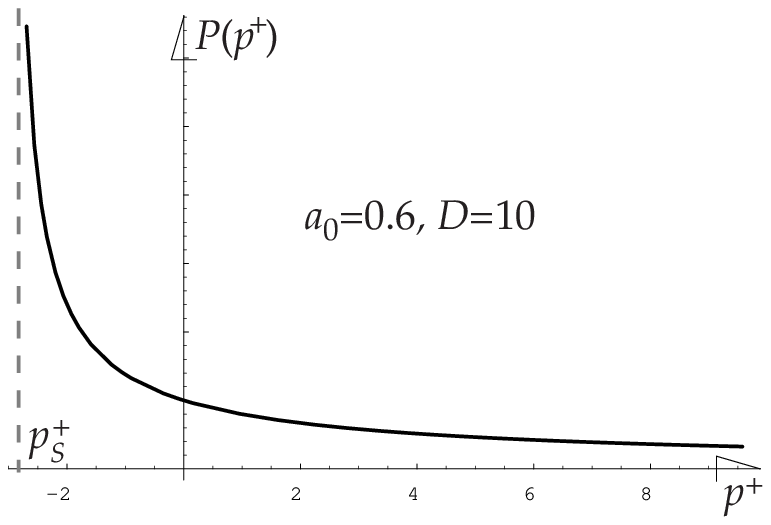}
       \hfill~}
       \caption{A plot of the distance function $P(p^+)$, for the source-less
       case, as a function of the proper distance, $p^+$. Note that
       $P(p^-)=P(2\tilde p_0-p^+)$, so that $P(p^-)$ is a reflection of
       $P(p^+)$ and the two plots both equal 1 at $\tilde p_0$.}
       \Label{f:P(p)}
\end{figure}
Finally, we compute the deficit angle at infinity
\begin{equation}
     \Delta = 2\p\Big(1-\lim_{p\to\infty}{P(p)\over p}\Big) = 2\p~.
     \Label{e:Delta0}
\end{equation}
Therefore the geometry of the transverse space is cylindrical,
with a varying radius $P(p)$.
    When  $a_0>0$ the cylinder is pinched at $p=\infty$, \ie\ its
circumference, $2\p P(p)$, rather quickly diminishes to zero.
On the other end, $P(p)$ diverges at $p_{\rm s}^+$: the circumference becomes
infinitely big. Thus, somewhat counterintuitively, the proper distance,
$p^+$, in the `outside' region of the transverse space grows from the
circle of infinite circumference at $p_{\rm s}^+$, toward circles of
vanishing circumferences, at $p^+\to\infty$.

Similarly, for $a_0<0$ the cylinder is pinched at $p=-\infty$, where the
circumference shrinks to a point. Also, $P(p)$ diverges at $p_{\rm s}^-$,
rendering the circumference infinitely big. In contrast to the `outside'
region region, in the `inside,' the proper distance, $p^-$, grows
from the circles of vanishing circumference at $p^-\to-\infty$, toward
the circle of infinite circumference, at $p^-=p_{\rm s}^-$.

The (linear) curvature scalar, $R_\perp$, of the transverse space is:
\begin{equation}
     R_\perp~=~-{D-3\over D-2}\,a_0^2\,l^{-2}
                 [1+a_0\log(r)]^{-{D-1\over D-2}}~.
     \Label{e:SC.nosrc}
\end{equation}
The limit $a_0\to 0$ is clearly interesting, since then this (invariant)
curvature of the transverse space vanishes, and so the transverse space
becomes flat. This justifies our earlier conclusions involving the
weak-field linearized Einstein tensor $\Q_{\mu\nu}$ in
Eq.~(\ref{e:PsTnosrc}).
    In addition, it is straightforward to see that
\begin{equation}
     \lim_{a_0\to0} P(p )~=~l{\tilde\l}^{-{D-3\over D-1}}~,
\Label{e:CFa0.nosrc}
\end{equation}
\ie\ the transverse space attains the shape of a flat cylinder of
constant circumference $2\p l {\tilde\l}^{-{D-3\over D-1}}$. This result
finally gives a transparent physical interpretation of this constant.
    Furthermore, from the expression for $p _S$ we see that the naked
singularity is pushed to $p_{\rm s}^+\to-\infty$ for $a_0\to +0$
($p_{\rm s}^-\to+\infty$ for $a_0\to +0$). Thus,
\begin{equation}
     \lim_{a_0\to0+}p^+\in(-\infty,+\infty)~, \quad\hbox{and}\quad
     \lim_{a_0\to0-}p^-\in(-\infty,+\infty)~.
\end{equation}
That is, both the `outside' and the `inside' region now extend over the
whole cylinder.

Following the standard procedure for T-duality (see e.g~\cite{clifford}),
we dualize on the $\q$ coordinate.
The only element in the T-dual metric that is different from the
original one~(\ref{e:metric}) is 
\begin{equation}
     \tilde{g}_{\q\q} ~\define~{1\over
g_{\q\q}}~=~{{\tilde l}^2\over\a'} [1+a_0\log(r)]^{D-3\over D-2}~.
\end{equation}
Here, $\tilde{l}=\a'\,l^{-1}$.
Of course, the dilaton now becomes non-constant:
\begin{equation}
     e^{2\tilde\F}~=~{{\tilde l}^2\over \a'}
g_s^2[1+a_0\log(r)]^{D-3\over (D-2)}~.
\end{equation}
Very near the naked singularity where $r\sim\rs$,
$e^{2\tilde\F}\sim g_s^2{{\tilde{l}}^2\over \a'}(r-\rs)$.

In this dual picture, all the previous results,
from~(\ref{e:dpnosrc}) to just before~(\ref{e:CF.nosrc}) still apply, but
Eq.~(\ref{e:CF.nosrc}) is now substituted by
\begin{equation}
     \tilde{P}(p)~=~\tilde{l}\,
     \Big[\tilde\l+{a_0\over2l}{D-1\over D-2}(p-\tilde p_0)\Big]
                ^{{D-3\over D-1}}.
     \Label{e:tCF.nosrc}
\end{equation}
Note that at the naked singularity,
$p_{\rm s}=\tilde p_0-{2l\tilde\l\over a_0}{D-2\over D-1}$ and
$\tilde{P}(p_{\rm s})=0$. On the other end, $P(+\infty)=\infty$ for
$a_0>0$, and
$P(-\infty)=\infty$ for $a_0<0$.

We again compute the deficit angle at infinity
\begin{equation}
     \Delta = 2\p\Big(1-\lim_{p\to\infty}{\tilde{P}(p)\over p}\Big) = 2\p~.
     \Label{e:tDelta0}
\end{equation}
Therefore the geometry of the transverse space is still cylindrical,
with a varying radius $\tilde{P}(p)$.
    This time, however, for both $a_0>0$ and $a_0<0$ the cylinder is pinched
at $p=p_{\rm s}^\pm$, \ie\ its circumference, $2\p\tilde{P}(p)$, rather
quickly diminishes to zero near the naked singularity. On the other end,
$\tilde{P}(p)$ diverges as $p^\pm\to\pm\infty$: the circumference becomes
infinitely big. This time, the proper distance, $p^+$, in the `outside'
region of the transverse space grows from the circle of vanishing
circumference at $p_{\rm s}^+$, toward circles of unbounded circumferences,
at $p^+\to\infty$.
    Again in contrast to the `outside'  region, the `inside' proper
distance, $p^-$, grows from the circles of unbounded circumference at
$p^-\to-\infty$, toward the circle of vanishing circumference, at
$p^-=p_{\rm s}^-$. Qualitatively, this image (with the `inside' and
`outside' regions patched at the naked singularity) is
a two-sheeted cone, see Fig.~\ref{f:LantJigg}.
We may thus understand the configurations
discussed here and in Ref.~\cite{bhmone} as the $\q$-duals of a
two-sheeted cone.

\begin{figure}[ht]
       \framebox{\epsfxsize=160mm%
       \hfill~% Here be picture.
       \epsfbox{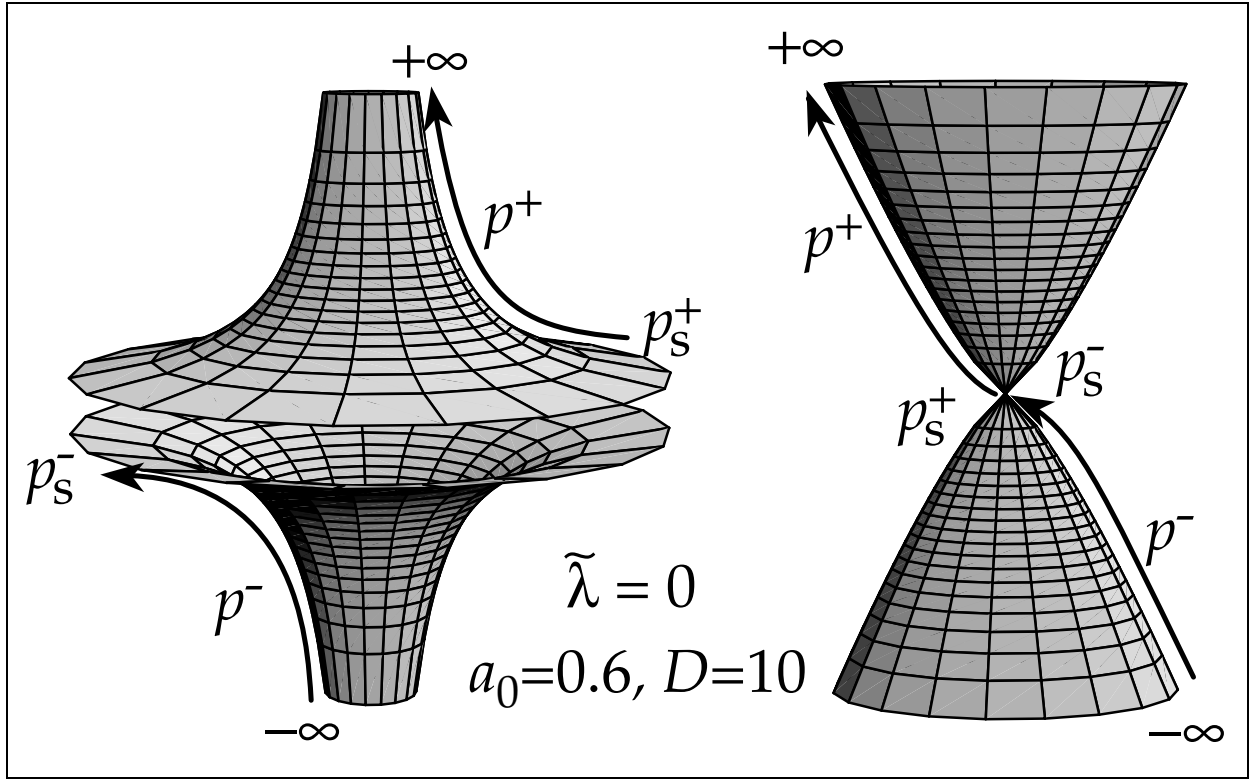}
       \hfill~}
       \caption{A plot of the two regions of the transverse space (left) and
         their $\q$-dual (right). The plot on the right is clipped at finite
         radius (given by $P(p^\pm)$; the two regions join at the naked
         singularity, at $P(p_{\rm s}^+)=\infty=P(p_{\rm s}^-)$.
         For $\tilde{\l}<0$, the two regions, and also their duals, separate;
         for $\tilde{\l}>0$ they intersect at a finite value of $P(p^\pm)$,
         \ie\ $\tilde{P}(p^\pm)$.}
       \Label{f:LantJigg}
\end{figure}

    In addition, it is straightforward to see that in the $a_0\to 0$ limit,
\begin{equation}
     \lim_{a_0\to0} \tilde{P}(p )~=~\tilde{l}{\tilde\l}^{{D-3\over D-1}}~,
     \Label{e:tCFa0.nosrc}
\end{equation}
\ie\ the transverse space attains the shape of a flat cylinder of
constant circumference $2\p l{\tilde\l}^{{D-3\over D-1}}$. In the radial
direction, we still have
\begin{equation}
     \lim_{a_0\to0+}p^+\in(-\infty,+\infty)~, \quad\hbox{and}\quad
     \lim_{a_0\to0-}p^-\in(-\infty,+\infty)~.
\end{equation}
That is, both the `outside' and the `inside' region now extend over the
whole cylinder.

\subsubsection{The case with sources}
Next we re-introduce sources in the Einstein equations.
The transverse metric is then
\begin{equation}
    \rd s_{\perp}^2 = l^2[1+a_0 \log(r)]^{-(D-3)\over(D-2)}
     r^{-2-\xi[2+a_0\log(r)]} (\rd r^2 + r^2\rd\theta^2).
\end{equation}
We repeat the analysis above and now find that
\begin{equation}
     p-p_0~=~{l\over2a_0}\,e^{\xi\over2a_0}\,
             \Big({2a_0\over\xi}\Big)^{D-1\over4(D-2)}\,
             \int_{t(r_0)}^{t(r)}\rd t~ t^{{D-1\over 4(D-2)}-1} e^{-t}~,
     \Label{e:PropDist}
\end{equation}
where we introduced $t\define{\xi\over2a_0}[1+a_0\log(r)]^2$.
The integral on the right hand side produces the incomplete
$\gamma$-function:
\begin{equation}
     p~=~p_0 + {l\over2a_0}\,e^{\xi\over2a_0}\,
         \Big({2a_0\over\xi}\Big)^{D-1\over4(D-2)}\,\bigg[\,
    \gamma\Big({D-1\over4(D-2)};{\xi\over2a_0}[1+a_0\log(r)]^2\Big)
    -\l\,\bigg]~.
    \Label{e:PD.src}
\end{equation}
As before, we have restricted $r$ to the physically permissible region,
$[\rs^+,+\infty)$ when $a_0>0$ and
$(0,\rs^-]$ when $a_0<0$. Therefore, with $r_0\in[\rs^+,+\infty)$,
\begin{equation}
     \l~\define~
         \gamma\Big({D-1\over4(D-2)};{\xi\over2a_0}[1+a_0\log(r_0)]^2\Big)
     \quad\To\quad \l\geq0~.   \Label{e:lambda}
\end{equation}

The resulting  expression is impossible to invert in closed form, and so
we are unable to express $P(p)$ in closed form, although it is of course
readily given in terms of the auxiliary (flat) $r$ coordinate:
\begin{equation}
    P(p(r))=l\,[1+a_0 \log(r)]^{-(D-3)\over 2(D-2)}
           r^{-\xi[1+{1\over2}a_0\log(r)]}.
           \Label{e:CF.src}
\end{equation}
Also, Fig.~\ref{f:P(p)xi} presents a plot, for three sample values of
$\xi$, of $P(p^+(r))$ {\it vs.\/} $p^+(r)$.
\begin{figure}[ht]
       \framebox{\epsfxsize=160mm%
       \hfill~%Here be picture.
       \epsfbox{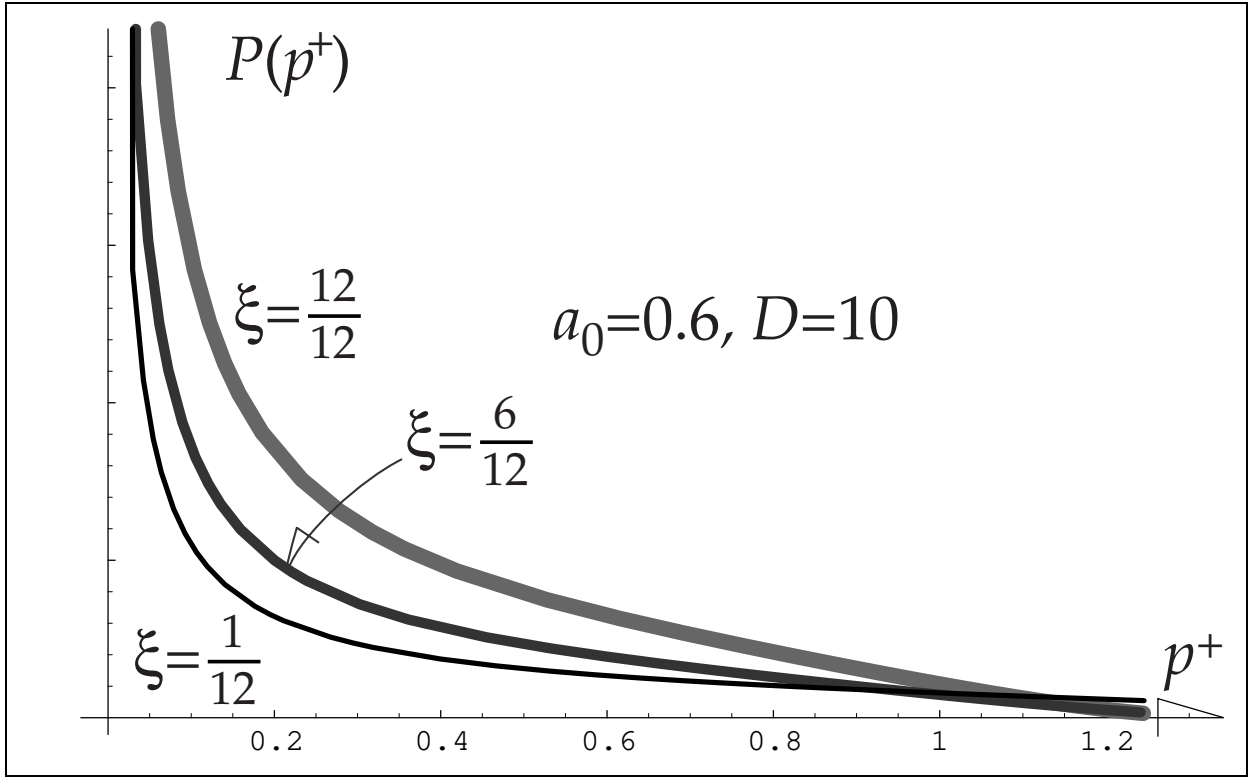}
       \hfill~}
       \caption{A (parametric) plot of the distance function $P(p^+(r))$ as a
       function of the proper distance, $p^+(r)$, plotted for the case with
       sources, for a few select values of $\xi$.}
       \Label{f:P(p)xi}
\end{figure}
However, note that $P (p)=0$ now occurs for finite $p$,
\begin{equation}
      p_{max}=p_0+{l\over2a_0}\,e^{\xi\over2a_0}\,
               \Big({2a_0\over\xi}\Big)^{D-1\over4(D-2)}
                \Gamma({D-1\over 4(D-2)})~,
\end{equation}
which has the correct limit $p_{max}\to\infty$ when $\xi\to0$, for the
source-less case. Other than that, $P(p^+)$ behaves the same for all
$0\leq\xi\leq1$.

The (linear) curvature scalar, $R_\perp$, of the transverse space now
becomes:
\begin{equation}
     R_\perp~=~-\Big[{D-3\over D-2}+2(\xi a_0^{-1})[1+a_0\log(r)]^2\Big]
     \,a_0^2\,l^{-2}[1+a_0\log(r)]^{-{D-1\over D-2}}\,
     r^{+\xi[2+a_0\log(r)]}~.
     \Label{e:SC.src}
\end{equation}
There are two interesting limits we can consider: $\xi\to0$, to
recover the source-less case, and $a_0\to 0$ (while keeping $\xi$
constant), which renders the transverse space flat.

Indeed, in the limit $\xi\to0$, the scalar curvature~(\ref{e:SC.src}) does
agree with the source-less result~(\ref{e:SC.nosrc}), as expected. Note
also that both of these curvatures vanish as $a_0\to0$, and $\xi$ is held
constant in~(\ref{e:SC.src}). As in the source-less case, the naked
singularity is gone in the $a_0\to0$ limit.

Let us now turn to discuss the $a_0\to0$ limit. In particular, we wish to
determine how the presence of sources ($\xi\neq0$) affects the dependence
of the length of the circumference, $2\p P(p)$, on the proper distance,
$p$. Although there is no expression $P(p)$ in closed form for arbitrary
$a_0$, in the $a_0\to0$ limit, the expression for $P(p)$ simplifies to
\begin{equation}
     P(r)|_{a_0=0} ~=~ r^{- \xi}~. \Label{e:CFa0.src}
\end{equation}
Furthermore, when $a_0\to0$ the expression for $p(r)$ simplifies into
\begin{equation}
       p-p_0~=~-{l\over \xi} r^{-\xi}~.
     \Label{e:PropDist0}
\end{equation}
Hence, Eq.~(\ref{e:CFa0.src}) becomes
\begin{equation}
     P(p)|_{a_0=0} ~=~l\xi(p_0-p) . \Label{e:CFap0.src}
\end{equation}
This is to be compared with the result~(\ref{e:CFa0.nosrc}) for the
source-less case. For positive $\xi$, and so also positive $a_0\to0+$
(recall the relations~(\ref{e:signa0xi})) we are in the `outside' region,
and Eq.~(\ref{e:CFa0.src}) clearly indicates that the circumference of the
cylindrical transverse space, $2\p P(p^+)\sim 2\p r^{-|\xi|}$ is shrinking
with a growing $p^+$; thus, the transverse space is pinching at
$p^+\to\infty$.

When $\xi$ is negative, $a_0\to0-$, and we are in the `inside'
region. Recall that now the proper distance, $p^-\to-\infty$ when $r\to0$.
The circumference, $2\p P(p^-)\sim2\p r^{|\xi|}$, of the cylindrical
transverse space again pinches at $p^-\to-\infty$. The two regions of the
transverse space thus turn out to have  very similar geometries.

It is now straightforward to compute the deficit angle:
\begin{equation}
     \Delta=2\p\Big(1 - \lim_{p\to\infty} {P(p)\over p}\Big) = 2\p(1+\xi)~.
\end{equation}

Thus the choice $a_0\geq 0$, in which case $\xi>0$, gives us the standard
$D7$-branes in the limit $a_0\to 0$, while $a_0\leq 0$ formally implies
that the number of branes is negative.

However, recall the special property of the codimension-2 solutions, that
the sources of Coulomb potentials are located at both $r=0$ and
$r=\infty$. To check this for the case $a_0\to 0-$ (for $a_0\leq 0$), we
change coordinates $r\to{1\over r}$. The distance function is then given
by, for $a_0<0$,
\begin{equation}
     P(p)|_{a_0=0} ~=~l|\xi|(p_0-p)~, \Label{e:CFap0xi.src}
\end{equation}
and hence the deficit angle becomes
\begin{equation}
     \Delta= 2\pi (1+|\xi|)~.
\end{equation}
The conclusion is then that for $a_0\to 0+$ the 7-branes are located at
$r=0$ while for $a_0\to 0-$ the 7-branes are located at
$r=\infty$.
As we will see in section~4 this is consistent with the resolution of the
naked singularity when the 7-brane is wrapped on a $K3$.

\subsection{Supersymmetric Limit and F-theory}
\Label{s:SuSy}
\subsubsection{Supersymmetry}
\Label{s:SuSy1}
With the above identification of our brane solution we now turn to
verify that the $a_0\to 0$ limit indeed gives rise to a supersymmetric
vacuum.

In order to establish that there exists a supersymmetric limit of  our
solution let us study the harmonic function $Z_7(r)$ which figures in the
expression for the metric~(\ref{e:A}-\ref{e:Bone}).
    For the metric~(\ref{e:metric}), the condition for having a Killing spinor
(\ie\ the supersymmetry variation of the gravitino) is that the first warp
factor, $e^{2A}$, is constant\footnote{For codimensions other than 2,
there is an additional condition that relates $A'(r)$ and $\F'(r)$. In
codimension-2, however, this condition becomes trivial for $A'(r)=0$.} (see
eg.~\cite{stelle}). That is, $A'(r)$ is the obstruction for supersymmetry.
From
\begin{eqnarray}
     e^{2A(r)}~&=&~ [1+a_0\log(r)]^{2\over D-2} ~\define~ Z_7^{2\over
D-2}~,\nn\\
     A'(r)&=&{1\over (D-2)}{a_0\over r(1+a_0\log(r))}~,
\end{eqnarray}
   we have $Z_7(r) \to 1$ and $A'(r)\to0$ when $a_0\to 0$. Thus
the obstruction to supersymmetry is lifted.
    While the first Chern class of the spacetime does vanish in both of the
other two limits, $\xi\to0$ and $l\to\infty$ (see
section~\ref{s:SpTmetric}), they do not affect the (non)vanishing of the
supersymmetry obstruction, $A'(r)$.
This unambiguously identifies $a_0\to0$
as the supersymmetric limit.

As further evidence let us now consider whether the solution saturates the
BPS bound:
\begin{equation}
      E~\geq~\Delta~. \Label{e:BPSBound}
\end{equation}
   In section~\ref{s:TrSpace}
we found that there exist two distinct contributions to the
deficit angle, $\Delta$, when $\xi\neq 0$. There is
on one hand the purely gravitational, or background part, $\Delta_0=2\p$,
from Eq.~(\ref{e:Delta0}). In addition, we have $\Delta_\xi=2\p|\xi|$ from
the stress-tensor,
$T_{\mu\nu}$. Thus, the total deficit angle is then
\begin{equation}
\Delta~=~\Delta_0~+~\Delta_\xi~=~2\p(1+|\xi|)~.
\end{equation}

We now turn to the tension, $E$. While it was possible to identify the
gravitational and stress tensor contributions to the deficit angle
separately, we only have a means of computing the tension due to the
non-zero $T_{\mu\nu}$.
The tension defined by~(\ref{e:Etension}) is
\begin{eqnarray}
     E_\xi&=&2\pi {\w^2\over 8a_0}
           \int_{\rs^+}^\infty{\rd r\over r} 2a_0[1+a_0 \log(r)]
      ~=~2\pi\xi \int_0^\infty \rd\rho~ 2a^2_0\rho~,\nn\\
      &=&2\pi\xi\int_0^{1/a_0} \rd\rho~ 2a^2_0\rho
         ~+~2\pi\xi\int_{1/a_0}^\infty \rd\rho~ 2a^2_0\rho~,\nn\\
      &\to&2\pi \xi~,\quad\hbox{when}~a_0\to0~,  \Label{e:Etensionsusy}
\end{eqnarray}
since the second integral vanishes in the $a_0\to0$ limit\footnote{The
evaluation of the integral~(\ref{e:Etensionsusy}) in the limit $a_0\to0$
is rather delicate: if the limit is taken prior to integrating, the result
would seem to vanish. Yet, the integral itself diverges and so does the
tension for any finite $a_0$. Therefore, the tension becomes an expression
of the form $0{\cdot}\infty$ in the $a_0\to0$ limit, and is precisely
evaluated by means of introducing the intermediate $1/a_0$ limit in the
second line of Eqs.~(\ref{e:Etensionsusy}).}.
We may now calculate $E_{-|\xi|}$ by applying the inversion
symmetry~(\ref{e:InvSymm}), and find that $E_{-|\xi|}=2\p|\xi|$,
so that in general we have
\begin{equation}
     E_\xi~=~2\p|\xi|~. \Label{e:TensionXi}
\end{equation}

    Thus, in this limit, the bound~(\ref{e:BPSBound}) is satisfied in the
modified version:
\begin{equation}
     E_\xi~\geq~\Delta_\xi~.
\end{equation}
We note that the tension density, \ie\ the integrand in
Eqs.~(\ref{e:Etensionsusy}), is supported either at $r=0$ (for $a_0\to0+$)
or at $r=\infty$ (for $a_0\to0-$).  Hence, the brane solution is now
localized, either at the origin or at infinity of  the transverse space,
as is familiar from the supersymmetric $D7$-brane solution~\cite{vafa}.

\subsubsection{Possible F-theory interpretation}
\Label{s:SuSy2}
We now explore a possible F-theory interpretation to this limit of our
solution along the lines of Sen's construction of constant-$\t$ solutions
in F-theory~\cite{sen}.
These solutions are classified in terms of the so called $j$-function,
\begin{equation}
      j(\tau(z))= f^3/\delta(f,g)~,
\end{equation}
where $f,g,\delta=4 f^3 + 27 g^2$ specify the elliptic fiber,
\begin{equation}
      u^2 = v^3 + f(z) v + g(z)
\end{equation}
as a function of the location $z$ in the plane perpendicular to
the brane. Here $f(z)$ and $g(z)$ are homogeneous
functions of degree 8 and 12 respectively.
If $j$ is constant, so is then $\tau$ --- and this is precisely
the case in the $a_0\to 0$ limit of our solutions.

To be more explicit, let us consider the source-free ($\xi=0$) case in
the $a_0\to 0$. From our previous discussion  this can be
interpreted as a collection of 12 $D7$-branes located at the origin,
$r=0$. However, as
$\tau$ is constant the $D7$ branes have to be configured such that the
total RR-charge is zero.

In order to reproduce this result from F-theory we choose
\begin{equation}
      f(z)=\alpha\prod_{i=1,2}(z-z_i)^2 \tilde f_4(z)\,,\quad
      g(z)=\prod_{i=1,2}(z-z_i)^3 \tilde g_6(z)\,,
\end{equation}
which leads to $j(z)\sim \alpha^3$. At each of $z=z_i$ there is a $D_4$
singularity, whose monodromy is $-\Ione$. By letting the two
singularities coalesce, the total monodromy is $\Ione$. Thus, we have
found agreement with our solution.

Here, the modulus $\tau$ is constant because of the fact that the negative
RR charge of  the orientifold plane is canceled by the $D7$ branes sitting
at the location of the orientifold planes. In general, however, this
argument fails, as orientifold interpretations exist only for certain
values of $\tau$~\cite{sen,dasgupta}. For a fixed but arbitrary
$\tau$, there does not exist any known orientifold interpretation.

Let us now turn to the $\xi\neq0$ case. The analysis in
section~\ref{s:SuSy1} shows that for $\xi<0$ (and so also $a_0\leq0$) the
branes are localized at $r=\infty$. This cluster of branes may then be
analyzied separately from the `background' cluster at $r=0$. In the
supersymmetric limit, $a_0\to0$, varying $\xi\in[0,-1]$ corresponds to
adding $12|\xi|$ branes at $r=\infty$. 
%%%(In particular when $|\xi|=1$ the transverse space is $S^2$.) 
Such configurations may be analyzed
much the same as was done above for the `background' cluster, at $r=0$.
    In contrast, $\xi>0$ (and so also $a_0\geq0$) the branes are localized at
$r=0$. This cluster of branes then overlaps with the `background' cluster,
and increases the degree of the singularity at $r=0$. To the best of our
knowledge, such highly degenerate configurations have not been discussed in
the literature.

Finally we briefly discuss the other solution, $\tau_{II}$. Our analysis
of the harmonic function $Z_7(r)$ when $a_0\to 0$
goes through as before.
However, there is a drastic difference compared to $\tau_I$
in the supersymmetric limit,
\begin{equation}
      \tau_{II}\to -\frac{n}{2\pi} + i \infty~.
\Label{e:tauIIlimit}
\end{equation}
This is indeed the correct behavior for a collection of $n$ $D7$-branes
located at $r=\infty$ as discussed
in the context of F-theory~\cite{vafa,zwiebach}.

\section{Probing the Solution}

In this section we continue to study spacetime properties of our
background. In the supersymmetric case the background geometry of
$N$ $D7$ branes can be nicely understood by sending a test probe (for
example, another $D7$ brane) from infinity. The physics of the brane probe
is described in terms of its effective mass and the effective potential
the test brane ``feels'' in the background it probes~\cite{clifford}.

Our background is non-supersymmetric in general. Yet we have shown that in
a particular limit we get a supersymmetric background of a certain
collection of $D7$ branes. It is natural thus to probe our solution with
a test $D7$ brane. However, we should keep in mind that although one
in principle can probe any background, whether it is made of
$Dp$-branes or not, the resulting physics will in general depend on the
type of brane probe. In particular, if a background is a near horizon
limit of some set of $Dp$ branes the most natural probe is a $Dp$ brane. If
a $Dp'$ brane is used instead this result will have to be consistent with
that of the $Dp$ brane probe analysis~\cite{cliffordII}.

The dynamics of an ideal D7-brane probe\footnote{We emphasize that the
brane probes we discuss are ideal in the sense that their presence is
assumed not to affect the geometry of the spacetime, and have no
back-reaction with the gravitational field probed.}
 is governed by the standard Born-Infeld
action~\cite{BI,joed}:
\begin{eqnarray}
       S_7&=& \int\rd^8x~\cL_7[G^s_{ab},\F,C_8]~, \nn\\
       \cL_7&=&-\mu_7 e^{-\Phi} \sqrt{- \det G^s_{ab}}
       + \mu_7 C_8~. \Label{e:BI}
\end{eqnarray}
Here $G^s_{ab}$ is the metric on the brane induced from the
background string frame metric by embedding the brane coordinates along
the spacetime ones. $C_8$ is the $8$-form potential whose field strength
is dual to $F_1 = dC_0$, where $C_0$ is the axion. We use the general
notation~\cite{clifford}
\begin{equation}
     \mu_p ~\define~ 2\p(2\p\sqrt{\a'})^{-(p+1)}
\end{equation}
for the brane tensions.
    All our configurations have $B_{\mu\nu}=0$, which simplifies the
Born-Infeld action~\cite{BI}. For notational simplicity, we
focus on the `outside' branch, where $r\in[\rs^+,\infty)$. The analysis
for the `inside' branch, where $r\in(0,\rs^-]$, is analogous.

\subsection{The $T_{\mu\nu}=0$ case}
\Label{s:PrT=const}
Since $\tau=\const$, so are the $C_0$ and $C_8$ potentials. Similarly,
the dilaton is also constant, $e^{\F}=g_s$.
    The induced string frame metric on the brane probe is~\cite{clifford}
\begin{equation}
     [G^s_{ab}]~=~e^{\F/2}g_s^{-1/2}\hbox{\rm diag}
       [\,(e^{2B}v^2{-}e^{2A})\,,\,\underbrace{e^{2A},\cdots,e^{2A}}_{7}\,]~,
       \Label{e:IndBrMet}
\end{equation}
where the factor of $e^{2\F}$ comes from the relation between the
Einstein and string frame metrics, $G_{ab}^s=e^{\F/2} G_{ab}^E$, and
the factor of $g_s^{-1/2}$ comes from the rescaling of the original
action~(\ref{e:effaction}).
The square root of the determinant of this metric is
\begin{eqnarray}
     \sqrt{- \det[G^s_{ab}]} &=& e^{2\F} g_s^{-2}
          Z_7(r)\sqrt{1-l^2\,Z_7(r)^{-9/8}r^{-2} v^2}~,\Label{e:sqrt}\\[1mm]
      &=& e^{2\F} g_s^{-2}
Z_7(r)\Big(1-\inv2 l^2\,Z_7(r)^{-9/8}r^{-2} v^2 ~+~O(v^4)\Big)~.
\Label{e:BIactionzero}
\end{eqnarray}
The expansion is in powers of $v^2=v_r^2 + r^2 v_\theta^2$, the square of
the velocity of the brane\footnote{By working in so called static gauge,
the brane coordinates can be aligned with the background. The induced
metric carries a time dependence reflecting how the brane moves in the
transverse direction.} in the transversal space. As usual, we may assume
that the velocity of the brane is small, so that we can neglect terms of
$O(v^4)$; we will return to this point shortly.

Inserting Eq.~(\ref{e:BIactionzero})
into~(\ref{e:BI}) and truncating at $O(v^2)$,
we get the following effective Lagrangian density for the test probe,
\begin{equation}
      \cL_7 ~\approx~ \mu_7[C_8- g^{-1}_s Z_7(r)]
      + \inv2 \mu_7 g^{-1}_s l^2 r^{-2}Z_7(r)^{-1/8} v^2~.
      \Label{e:LTzero}
\end{equation}
Upon integration over unit 7D-volume along the brane, this is simply a
non-relativistic Lagrangian for a particle  with potential and
kinetic energies given by
\eqn
      V(r) &=& \mu_7 V_{D7}[g^{-1}_s Z_7(r)-C_8]~,\Label{e:V(r)}\\
      T(r) &=& \inv2 m(r) v^2~,\Label{e:T(r)}\\
      m(r)&\define&\mu_7V_{D7} g^{-1}_s l^2 Z_7(r)^{-1/8} r^{-2}~,
      \Label{e:m(r)}
\enn
where $V_{D7}\define\int\rd^7x$ is the spatial 7D-volume of the brane
probe.
    Although the effective mass~(\ref{e:m(r)}) of the brane probe varies with
$r$, it remains positive over the physically permissible region,
$r\in[\rs^+,\infty)$ for $a_0>0$ and $r\in(\infty,\rs^-]$ for $a_0<0$.
However, at the naked singularity $m(r)$ diverges. Also unlike in the
supersymmetric case, neither the effective potential for the brane probe
nor its velocity are constant in the present configurations.

It behooves us then to use the exact expression~(\ref{e:sqrt}), rather than
its truncation~(\ref{e:BIactionzero}).
Then, the exact Lagrangian density~(\ref{e:BI})
becomes:
\begin{equation}
  \cL_7~=~\mu_7[C_8-g_s^{-1}Z_7(r)\sqrt{1-l^2\,Z_7(r)^{-9/8}r^{-2}v^2}]~,
   \Label{e:BIdet}
\end{equation}
from which we define, performing the Legendre transform, the
momentum (assuming $\dot\q=0$, for simplicity) and the total energy
(density):
\begin{eqnarray}
  \wp&=&{\vd\cL_7\over\vd v}~=~
   {\mu_7g_s^{-1}l^2r^{-2}Z_7^{-1/8}\over
     \sqrt{1-l^2\,Z_7(r)^{-9/8}r^{-2}v^2}}\,v~,
   \Label{e:rMom} \\
  \cE&=& \wp\,v-\cL_7~=~
   {\mu_7g_s^{-1}Z_7\over
          \sqrt{1-l^2\,Z_7(r)^{-9/8}r^{-2}v^2}}-\mu_7C_8~.
   \Label{e:totE}
\end{eqnarray}
For future reference, we define the `relativistic' counterparts
of~(\ref{e:V(r)}) and~(\ref{e:T(r)}):
\begin{eqnarray}
  \cV_7&=&\cE\Big|_{v=0}~=~\mu_7g_s^{-1}Z_7-\mu_7C_8~,\Label{e:rV(r)}\\
  \cT_7&=&\cE-\cV_7~=~\mu_7g_s^{-1}Z_7\bigg[
   {1\over\sqrt{1-l^2\,Z_7(r)^{-9/8}r^{-2}v^2}}-1\bigg]~,
    \Label{e:rT(r)}
\end{eqnarray}

From the form of the total energy (density), we may interpret
\begin{equation}
  c_{\rm eff}~=~rl^{-1}\,Z_7(r)^{9/16} \Label{e:cEff}
\end{equation}
as the effective speed of light for the brane probe approaching the naked
singularity, and note that $\lim_{r\to\rs^+}c_{\rm eff}=0$.

Next, we assume that the total energy (density), $\cE$, is
conserved\footnote{The conservation of total energy is reasonable, as the
brane probe is assumed to be ideal: it probes the spacetime geometry and
so interacts only with the effective potential. Thus, unlike a realistic
probe, it cannot lose energy through radiation or decay.} and may be used
to parametrize the velocity:
\begin{equation}
     v~=~{r\over l}Z_7(r)^{9/16}
      \sqrt{1-\Big({\mu_7g_s^{-1}Z_7(r)\over\cE+\mu_7C_8}\Big)^2}~.
      \Label{e:v(E)}
\end{equation}
For the velocity of the brane probe to be real (as appropriate for the
classical analysis), it must be that:
\begin{equation}
  Z_7(r)\geq0~\quad\hbox{and}\quad
  1\geq\Big({\mu_7g_s^{-1}Z_7(r)\over\cE+\mu_7C_8}\Big)^2~. \Label{e:conds}
\end{equation}
The first of these implies that $r\geq\rs^+=e^{-1/a_0}$, as was established
in section~\ref{s:SpTimeProps}.
 Taking the square-root of both sides of the second
condition~(\ref{e:conds}), we obtain the lower bound on the total energy
(density):
\begin{equation}
  \cE+\mu_7C_8~\geq~\mu_7g_s^{-1}Z_7(r)~,\quad\hbox{where}\quad
      Z_7(r)~\geq~0~. \Label{e:EOut}
\end{equation}
Solving this for $r$, we find the outer limit:
\begin{equation}
  r~\leq~\ro~\define~\exp
      \bigg\{a_0^{-1}\Big({\cE+\mu_7C_8\over g_s^{-1}\mu_7}-1\Big)\bigg\}~.
       \Label{e:rOut}
\end{equation}

From Eq.~(\ref{e:v(E)}) (see also the plot of $v(r)$ in
Fig.~\ref{f:limits}), as the brane probe approaches $\ro$ from `inside' or
$\rs^+$ from `outside,' its velocity drops to zero, and the brane probe
starts returning within the `window' $r\in[\rs^+,\ro]$. (By conservation of
energy, the brane probe will continue moving after hitting the naked
singularity.) Both $\rs^+$ and $\ro$ are classical turning points, and the
brane probe is being ricocheted between them.
    This result is in perfect accord with the result in the Appendix~A
of~\cite{bhmone}, from which it follows that a quantum particle probe
must remain in a scattered state although the potential is attractive.
Note though that the transverse volume available to the brane probe is
determined by size of the `window' $r\in[\rs^+,\ro]$; see
section~\ref{s:exphierarchy}, rather than $r\in[\rs^+,\infty]$.

One of the assumptions of the brane probe analysis is that the branes are
moving {\sl very slowly} in the given background. We are now in the
position to make this statement more precise for the present
configuration.
    By substituting the exact result~(\ref{e:v(E)}) into the Lagrangian
density~(\ref{e:BIdet}), it is easy to verify that the
square-root~(\ref{e:sqrt}), and so also the entire Born-Infeld action are
real. Note that this is equivalently ensures that the `lapse function'
along the brane probe, $G_{tt}=(e^{2B}{-}e^{2A})$ in the induced
metric~(\ref{e:IndBrMet}), remains negative. Thus, the spacetime of
the brane probe remains of the Lorentzian signature everywhere in the
physically permitted transverse space.
    Moreover, the Taylor expansion of the square-root~(\ref{e:sqrt})
always converges, owing to~(\ref{e:v(E)}), and also the $O(v^4)$ terms
are {\it always\/} smaller than the $O(v^2)$ ones:
\begin{equation}
      \inv2l^2\,Z_7(r)^{-9/8}r^{-2}v^2
    ~>~\inv8(l^2\,Z_7(r)^{-9/8}r^{-2}v^2)^2~.
\end{equation}
This is, owing to the exact result~(\ref{e:v(E)}), equivalent to
\begin{equation}
    3~>~-\Big({\mu_7g_s^{-1}Z_7(r)\over\cE+\mu_7C_8}\Big)^2~,
\end{equation}
which is trivially satisfied and restricts neither $r$ nor $\cE$. Hence
the description of a classical particle moving in a potential $V(r)$
and with kinetic energy $T(r)$ given by~(\ref{e:V(r)}--\ref{e:m(r)})
provides a consistent approximation when $r\in[\rs^+,\infty]$.

Note however that as $r\to\rs^+$, both the velocity of the
probe~(\ref{e:v(E)}) and the effective speed of light~(\ref{e:cEff}) go to
zero, in such a way that
\begin{equation}
  \Big({v\over c_{\rm eff}}\Big)~=~
   \sqrt{1-\Big({\mu_7g_s^{-1}Z_7(r)\over\cE+\mu_7C_8}\Big)^2}
    ~\to~1~.
\end{equation}
Furthermore, near $\rs^+$, all of the brane probe's energy becomes `kinetic'.
This suggests that neglecting the relativistic effects such as
gravitational back-reaction, perturbation of the background, {\it etc.\/}
is no longer warranted. To ensure that the `non-relativistic'
formulae~(\ref{e:V(r)}--\ref{e:m(r)}) are still reliable, we require that
the difference between the relativistic kinetic energy~(\ref{e:rT(r)}) and
the non-relativistic one~(\ref{e:T(r)}) be smaller than the latter:
\begin{equation}
  \cT_7(r)-T(r)V_{D7}^{-1} ~<~T(r)V_{D7}^{-1}~, \quad\hbox{\ie}\quad
  \cT_7(r)~<~2T(r)V_{D7}^{-1}~. \Label{e:Tbound}
\end{equation}
Using the exact result for the velocity~(\ref{e:v(E)}), this produces
the {\it upper\/} bound on the total energy (density):
\begin{equation}
  \cE+\mu_7C_8~\leq~\b\,\mu_7g_s^{-1}Z_7(r)~,\qquad
   \b~\define~\sqrt{2\over3-\sqrt5}\approx1.618~.
  \Label{e:EIn}
\end{equation}
In turn, this produces an {\it inner\/} limit on the radius:
\begin{equation}
  r~\geq~\ri~\define~\exp
   \bigg\{a_0^{-1}\Big({\cE+\mu_7C_8\over\b g_s^{-1}\mu_7}-1\Big)\bigg\}~.
       \Label{e:rIn}
\end{equation}

Now, we may keep reducing\footnote{The brane probe here is an ideal
object, the total energy density of which is simply a control; we do not
address the nature of the physical mechanism by which to decrease the
brane probe's energy. However, it is conceivable that a real probe would
lose energy through gravitational bremsschtrahlung, if nothing else.} the
total energy (density) of the brane probe towards the lower
limit~(\ref{e:EOut}), $\cE\to-\mu_7C_8$, in the hopes of decreasing
$\ri\to\rs^+$ so as to approach the naked singularity. This however also
reduces $\ro\to\rs^+$, and the allowed region, $[\ri,\ro]$, for the
brane probe becomes vanishingly small as it nears the naked singularity:
the (classical) brane probe is trapped by it.
Alternatively, $\ri\to\rs^+$ by taking $a_0\to 0$ while keeping
${\cE+\mu_7C_8\over\mu_7g_s^{-1}}>1$, as can be seen from
Eqs.~(\ref{e:rOut}) and~(\ref{e:rIn}).

The two limits, (\ref{e:rOut}) and~(\ref{e:rIn}), provide a region in
which the classical probe analysis based on the action~(\ref{e:BI}) is
limited:
\begin{equation}
     \rs^+~<~\ri ~\leq~ r ~\leq~ \ro~, \Label{e:rWindow}
\end{equation}
where the left-most inequality holds since $\cE{+}\mu_7C_8\geq0$, owing to
the `outer reality' condition~(\ref{e:EOut}).
As discussed above, the outer limit $\ro$ is a simple classical
turning point, while below $\ri$ the action~(\ref{e:BI}) no longer
suffices to describe the dynamics of the brane probe. The distance from the
naked singularity to these limits grows exponentially, but remains
finite for all finite energies. By comparison, there is no `outer' turning
point for higher codimension brane probes~\cite{clifford,BHM2a}: there the
velocity is constant and an outgoing brane probe would leave the
singularity forever. Obviously, this distinction happens because, unlike
its higher codimension analogues,
$Z_7(r)$ has no asymptotic region and in fact diverges at $r=\infty$.

The situation is illustrated in Fig.~\ref{f:limits}.
\begin{figure}[ht]
       \framebox{\epsfxsize=160mm%
       \hfill~%Here be picture.
       \epsfbox{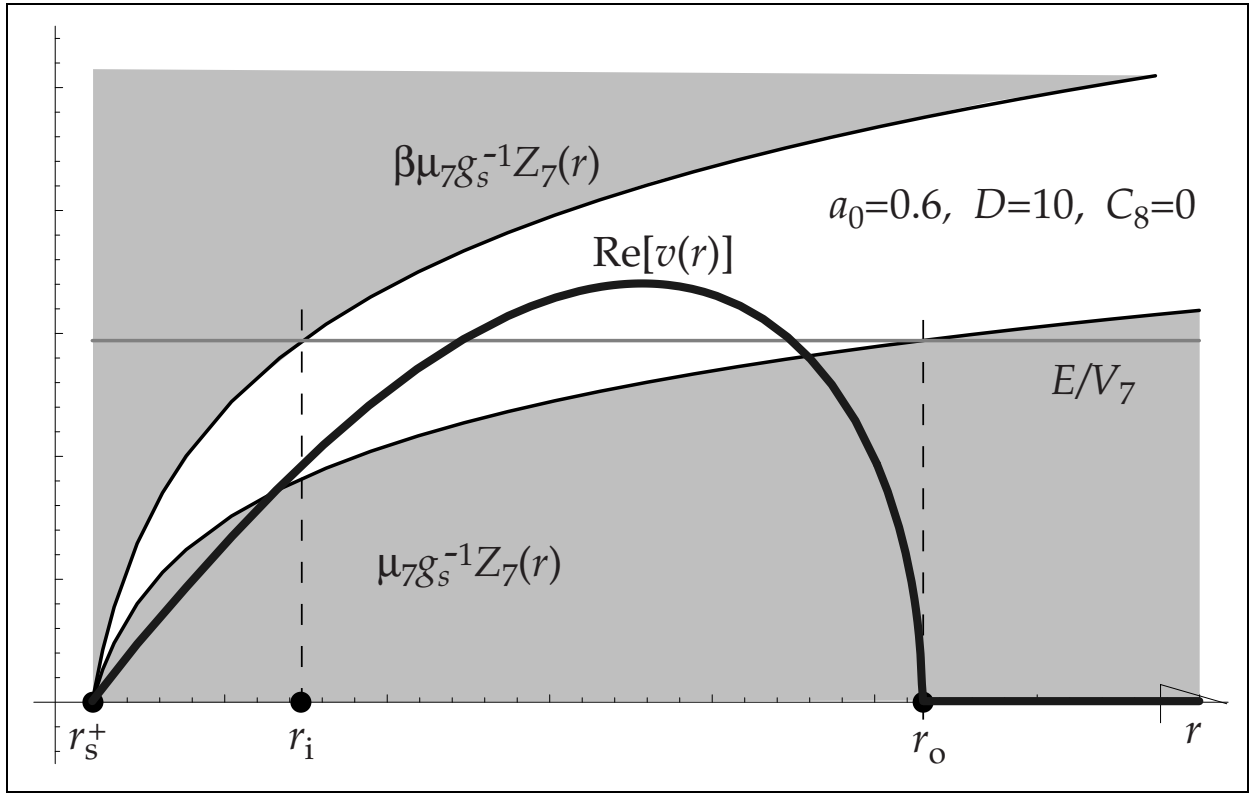}
       \hfill~}
       \caption{The `outer' limit curve (lower) and the `inner'
limit curve (upper); shaded areas are forbidden. The brane probe velocity,
$v(r)$ as defined in Eq.~(\ref{e:v(E)}), is plotted (grey). Although $v$
decreases as $r<\ri$, $v/c_{\rm eff}\approx0.786$ at $\ri$ and {\it
grows\/} as $r\to\rs^+$. Within $[\rs^+,\ri]$, the bound~(\ref{e:Tbound})
is violated; the brane probe based on the action~(\ref{e:BI}) is no longer
reliable.}
  \Label{f:limits}
\end{figure}
The radii where a given energy density level crosses the two limiting
curves are $\ro$ and $\ri$, respectively.
    In turn, Eqs.~(\ref{e:EOut}) and~(\ref{e:EIn}) can be combined to
produce the extended inequality
\begin{equation}
     \mu_7[g_s^{-1}Z_7(r)-C_8]~\leq~\cE
     ~\leq~\mu_7[\b g_s^{-1}Z_7(r)-C_8]~.
     \Label{e:Extended}
\end{equation}
Here both the upper and the lower limit depend on $r$, through $Z_7(r)$.
(While there does exist an analogous `band' of allowed brane probe energies
in the higher codimension case, its lower limit is constant~\cite{BHM2a}.)
    It is of course possible to choose the constant dual-axion field strength
so that the lower limit on the total energy of the brane probe would
vanish at any particular point of interest: $C_8=g_s^{-1}$ makes it vanish
in the surrogate asymptotic region $r\sim1$. Furthermore,
the width of the `band' of energies allowed by the
inequalities~(\ref{e:Extended}) is
\begin{equation}
     \Delta \cE~=~(\b{-}1)\mu_7g_s^{-1}Z_7(r)
      ~=~2(\b{-}1)\p(2\p\sqrt{\a'})^{-8}g_s^{-1}Z_7(r)~,
\end{equation}
and varies with $Z_7(r)$.

In the limit $a_0\to 0$, the inequalities~(\ref{e:Extended}) simplify to
\begin{equation}
     \mu_7[g_s^{-1}-C_8]~\leq~\cE~\leq~\mu_7[\b g_s^{-1}-C_8]~.
    \Label{e:SuSyWindow}
\end{equation}
Again, we may choose the constant dual-axion field strength,
$C_8\to g_s^{-1}$, whereupon the total energy of the brane probe is
limited to be $0\leq \cE\leq(\b{-}1)\mu_7g_s^{-1}$.
    With the total energy within these limits, we now find that
\begin{equation}
     \lim_{a_0\to0} \ro=+\infty~,\quad\hbox{and}\quad
     \lim_{a_0\to0} \ri=~0~=\lim_{a_0\to0} \rs^+~.
\Label{e:rlimit}
\end{equation}
That is, in the $a_0\to0$ limit, the probe analysis based on the
expansion~(\ref{e:BIactionzero}) is valid in the whole $r\in[0,\infty)$
region.

Next, $V(r)$ becomes a constant and the mass~(\ref{e:m(r)}) of the probe
moving around the cylinder is constant. In addition, the mass of the probe
moving in the radial direction, when measured in the proper distance, is
also constant. This is consistent with supersymmetry. It is
hence satisfying to see how the probe analysis reproduces this fact.

Finally, even the perhaps somewhat curious looking region of allowed
energies~(\ref{e:SuSyWindow}) can also be derived~\cite{BHM2a} from the
standard analysis of supersymmetric 7-branes~\cite{clifford}.

\subsection{The $T_{\mu\nu}\neq 0$ case}

\subsubsection{$\tau_I$}
This case resembles the source-free situation in that
$\Ree(\tau_I)=\const$ and hence both $C_{0,8}$ are constant although the
dilaton is now non-trivial
\begin{equation}
e^{2\F(\q)}=g_s^2 e^{-2\omega\theta}~.
\end{equation}
The probe analysis follows along the lines of section~\ref{s:PrT=const}.
  From Eq.~(\ref{e:BI}) and the expansion of the determinant of the induced
metric,
we find that the Lagrangian is given by
\begin{equation}
      \cL_7 = \mu_7[C_8- g^{-1}_s e^{-\w\q} Z_7(r)]
      + \inv2 \mu_7 g^{-1}_s  l^2 e^{-\w\q} r^{-2}
      Z_7(r)^{-1/8} e^{\xi a_0^{-1} - \xi a_0^{-1} Z_7^2(r)} v^2 +O(v^4)~.
      \Label{e:LTI}
\end{equation}
Upon integration over unit 7D-volume along the brane, this again is a
Lagrangian for a particle moving with potential and kinetic energies
\eqn
      V(r,\q) &=& \mu_7 V_{D7}[g^{-1}_s e^{-\w\q} Z_7(r)-C_8]\\
      T(r,\q) &=& \inv2 m(r,\q) v^2~,\\
      m(r,\q) &\define&\mu_7V_{D7} g^{-1}_s l^2 e^{-\w\q} Z_7(r)^{-1/8}
r^{-2}
                e^{\xi a_0^{-1} - \xi a_0^{-1} Z_7^2(r)}~.
      \Label{e:m(r)I}
\enn
In analogy with our previous discussion, when $T_{\mu\nu}=0$,
the `outer' and `inner'
reality conditions are
\begin{eqnarray}
     \cE+\mu_7C_8 &\geq&\mu_7g_s^{-1}e^{-\w\q}Z_7(r)~,\quad\hbox{where}\quad
      Z_7(r)~\geq~0~,
       \Label{e:EOutI}\\[1mm]
     \ie\quad r&\leq& \ro~\define~\exp
      \bigg\{a_0^{-1}\Big({\cE+\mu_7C_8\over
             g_s^{-1}e^{-\w\q}\mu_7}-1\Big)\bigg\}~,
       \Label{e:rOutI}
\end{eqnarray}
and
\begin{eqnarray}
  \b\,\mu_7g_s^{-1}e^{-\w\q}Z_7(r)&\geq&\cE+\mu_7C_8~,
   \Label{e:EInI}\\[1mm]
  \ie\quad r&\geq&\ri~\define~\exp
   \bigg\{a_0^{-1}\Big({\cE+\mu_7C_8\over
             \b g_s^{-1}e^{-\w\q}\mu_7}-1\Big)\bigg\}~,
       \Label{e:rInI}
\end{eqnarray}
respectively.
As for the $T_{\mu\nu}=0$ case, from~(\ref{e:rOutI}) and~(\ref{e:rInI}), we
get a window of allowed values for $r$,
\begin{equation}
     \rs^+~<~\ri ~\leq~ r ~\leq~ \ro~. \Label{e:rWindowI}
\end{equation}
Note that the classical limits $\ri$ and $\ro$ now depend on $\q$.
Eqs.~(\ref{e:EOutI}) and (\ref{e:EInI}) then imply an
extended inequality analogous to~(\ref{e:Extended}),
\begin{equation}
     \mu_7[g_s^{-1}e^{-\w\q}Z_7(r)-C_8]~\leq~\cE
     ~\leq~\mu_7[\b g_s^{-1}e^{-\w\q}Z_7(r)-C_8]~.
     \Label{e:ExtendedI}
\end{equation}
Here both the upper and the lower limit depend on $r$, through $Z_7(r)$,
and $\q$ through the dilaton dependence, $e^{\F(\q)}$.
Owing to the $\q$-depenence of the limits, the limits on the brane probe total
energy can be simplified only in a fixed, albeit arbitrary direction $\q_a$.
Furthermore, the width of the
`band' of energies allowed by the inequalities~(\ref{e:ExtendedI}) is
\begin{equation}
     \Delta \cE~=~(\b{-}1)\mu_7g_s^{-1}e^{-\w\q}Z_7(r)
      ~=~2(\b{-}1)\p(2\p\sqrt{\a'})^{-8}g_s^{-1} e^{-\w\q} Z_7(r)~,
\end{equation}
and varies with $Z_7(r)$ and $e^{\w\q}$.

In the limit $a_0\to 0$, the inequalities~(\ref{e:ExtendedI}) simplify to
\begin{equation}
   \mu_7[g_s^{-1}-C_8]~
\leq~\cE~\leq~\mu_7[\b g_s^{-1}-C_8]~.
    \Label{e:SuSyWindowI}
\end{equation}
We may choose $C_8\to g_s^{-1}$, so the total energy of the brane probe
is limited to be
\begin{equation}
     0\leq \cE\leq(\b{-}1)\mu_7g_s^{-1}~\quad\hbox{for}\quad \q=\q_a~.
\end{equation}
    With the total energy within these limits, we now find that
\begin{equation}
     \lim_{a_0\to0} \ro=+\infty~,\quad\hbox{and}\quad
     \lim_{a_0\to0} \ri=~0~=\lim_{a_0\to0} \rs^+~.
\end{equation}
That is, in the $a_0\to0$ limit, the probe analysis based on the
expansion~(\ref{e:BIactionzero}) is valid in the whole $r\in[0,\infty)$
region.

As before $V(r)$ and the mass of the probe
moving around the cylinder become constant. Also, the mass of
the probe moving in the radial direction, when measured in the proper
distance, is constant. Both of these facts are again consistent with
supersymmetry.

\subsubsection{$\tau_{II}$}

    From the expression for $\tau_{II}$ we have
\eqn
C_8(r)&=&{\omega\over a_0} \tanh(\omega\pi) Z_7^2(r) C^{(0)}_8~,\\
e^{\F(\q)} &=& 2 g_s \tanh(\omega\pi) \cosh(\omega\theta)~.
\Label{e:phitauII}
\enn
The probe analysis follows along the lines of section~\ref{s:PrT=const}.
  From Eq.~(\ref{e:BI}) and the expansion of the determinant of the induced
metric,
we find that the Lagrangian is given by
\begin{equation}
      \cL_7 = \mu_7[C_8(r)- g_s^{-2}e^{\F(\q)} Z_7(r)]
      + \inv2 \mu_7   l^2  g_s^{-2}e^{\F(\q)}
      Z_7(r)^{-1/8} r^{-2-\xi(2+a_0\log(r))} v^2 +O(v^4)~.
      \Label{e:LTII}
\end{equation}
Upon integration over unit 7D-volume along the brane, the effective
potential and kinetic energies of the probe are
given by
\eqn
      V(r,\q) &=& \mu_7 V_{D7}[g_s^{-2}e^{\F(\q)} Z_7(r)-C_8(r)]\\
      T(r,\q) &=& \inv2 m(r,\q) v^2~,\\
      m(r,\q) &\define&\mu_7V_{D7}  l^2 g_s^{-2}e^{\F(\q)} Z_7(r)^{-1/8}
   r^{-2-\xi(2+a_0\log(r))}~.
      \Label{e:m(r)II}
\enn

As discussed in section~\ref{s:PrT=const} assuming
that the total energy, $E$,
is conserved, we find the following range of allowed energies
\begin{equation}
     \mu_7[g_s^{-2}e^{\F(\q)}Z_7(r)-C_8(r)]~\leq~\cE
     ~\leq~\mu_7[\b g_s^{-2}e^{\F(\q)}Z_7(r)-C_8(r)]~.
     \Label{e:ExtendedII}
\end{equation}
Here both the upper and the lower limit depend on $r$, through
$Z_7(r)$ and $C_8(r)$, and $\q$ through $e^{\F(\q)}$.
As before, (\ref{e:ExtendedII}) determines the `inner' and `outer'
limits, $\ri$ and $\ro$.

In contrast to the analysis in sections~3.1 and 3.2 in the limit $a_0\to 0$,
the inequalities~(\ref{e:ExtendedII})
enforce $\cE=-\mu_7C_8$ rather than a window of allowed energies.
The solutions for $\ri$ and $\ro$ can be obtained
in analogy with Eq.~(\ref{e:rlimit}),
\begin{equation}
     \lim_{a_0\to0} \ro=~\infty~,\quad\hbox{and}\quad
     \lim_{a_0\to0} \ri=~\infty~=\lim_{a_0\to0} \rs^+~.
\end{equation}
Thus, in the $a_0\to0$ limit, the probe analysis breaks down.

\section{Non-supersymmetric Resolution of Naked Singularities}

We found in the last section that useful information can be learned by
studying a non-supersymmetric background with a supersymmetric $Dp$
brane.
While this particular set-up is interesting in its own right, we now
turn to a potentially more exciting configuration in which the
background is wrapped on a compact four-dimensional manifold, such as $K3$.
 From a phenomenological point of view
we are eventually interested in a four-dimensional theory. On the
other hand, $K3$'s non-trivial curvature induces an effective negative
number of $D3$ branes. This leads to a situation which is analogous to
the \enh-mechanism~\cite{joep}, although in a non-supersymmetric
setting. While we will compute  the effect of the
wrapping on the exponential hierarchy, we will focus on the formal
aspects of this problem.

\subsection{The $T_{\mu\nu}=0$ case}
\Label{s:enhancon}
We start by looking at the solution compactified on a $T^4$. By T-duality,
the 7-brane is turned into a 3-brane, which however
is smeared along the $T^4$. The metric, dilaton and four-form
potential are obtained, by following the analysis of Ref.~\cite{clifford}:
\eqn
ds^2 &=& Z_3^{2/4} \eta_{ab} dx^a dx^b + V^{1/2}_{T^4} ds^2_{T^4} +
l^2 Z_3^{-5/8}  r^{-2}(dr^2 + r^2 d\theta^2)~,\\
e^{2\F} &=& g_s^{2} Z_3^{-1}~,\\
C_4&=&\const~,\\
Z_3&=& 1 + a_0 {(2\pi \a'^{1/2})^4\over V_{T^4}} \log(r)~.
\enn
Note that the
non-trivial dilaton compensates for the $Z_3^{-5/8}$ behavior rather
than $Z_3^{-3/4}$ as would have been expected for a 3-brane in $D=6$ from
our solution~(\ref{e:A}-\ref{e:Bone}).
In fact, the behavior of the dilaton is opposite to
that of the supersymmetric $D3$- and $D7$-branes. There the $D3$ brane
has a constant dilaton while the $D7$ brane has a logarithmic running.

In wrapping a supergravity background on a $K3$ there is a non-trivial
$R\wedge R$ coupling, absent for the $T^4$. This has the effect of
inducing a negative number of
3-branes~\cite{bsv}. We will assume that this is
the only non-trivial coupling between the $K3$ and background although
there is no supersymmetry to protect other couplings from
contributing~\footnote{This assumption is justified by the
self-consistency of the analysis conducted below; in the $a_0\to
0$ limit we recover the results of~\cite{joep}.}.
Therefore we obtain the following wrapped solution
\eqn
ds^2 &=& Z_3^{2/4} Z_7^{1/4}  \eta_{ab} dx^a dx^b +
V^{1/2}_{K3} Z_7^{1/4}ds^2_{K3} + \nn\ \\
&&l^2 Z_3^{-5/8} Z_7^{-7/8} r^{-2}(dr^2 + r^2 d\theta^2)~,\\
e^{2\F} &=& g_s^{2} Z_3^{-1}~, \Label{e:DilZ3}\\
C_4&=&\const, C_8=\const~,\\
Z_3&=& 1 - a_0 {(2\pi \a'^{1/2})^4\over V_{K3}} \log(r)~,
     \Label{e:Z3}\\
Z_7&=& 1 + a_0  \log(r)~,
\Label{e:wrapsoln}
\enn
where we have used the standard procedure for generating a
supergravity solution of intersecting branes~\cite{stelle}. We however wish
to comment on the form of $Z_3(r)$. Firstly, the $\log(r)$ behavior is due
the the 3-brane being smeared over the 4-dimensional $K3$, so that the true
transverse space is still 2-dimensional. In addition, we know that
$Z_3(r)\to1$ if $V_{K3}\to\infty$, \ie\ if the volume of $K3$ becomes
infinite and we have effectively uncompactified the spacetime. This
rescales the $a_0$ parameter, the precise form of which is determined by
dimensional analysis as shown in~(\ref{e:Z3}). Finally the relative sign
in $Z_3(r)$ is determined along the lines of the argument, given in
Ref.~\cite{bsv}, for the emergence of a negative number of 3-branes.

With the wrapped solution from~(\ref{e:wrapsoln}) at hand we turn to
the probe analysis. Because of the $R\wedge R$ coupling the probe
action has to be modified (see e.g.~\cite{clifford})
\eqn
S &=& - \int_{M} e^{-\F}(\mu_7 V_{K3}(r) - \mu_3)
\sqrt{- \det G^s_{ab}} +\nn\ \\
&&\mu_7 \int_{M\times K3} C_8 - \mu_3 \int_M C_4~.
\Label{e:BIwrap}
\enn
Here,
\begin{equation}
     V_{K3}(r)~=~V_{K3} \Big({Z_7\over Z_3}\Big)^{1/2}~,
     \Label{e:V(r)K3}
\end{equation}
is the volume of the $K3$ in the
string frame metric.
As in the unwrapped case $C_{4,8}$ are both constant.

In order to understand the important features of this expression let
us briefly recapitulate the salient features in the supersymmetric case;
for more details, see~\cite{joep}. First, for a set of supersymmetric
D7-branes wrapped on a $K3$, the potential seen by a D7-brane probe is
constant. Second, the kinetic term has a mass $m_{3,7}\sim (\mu_7 V_{K3}
Z_3-\mu_3 Z_7)$. However, while
the harmonic function for D7-brane is the same as in our case,
the interpretation of $a_0$ is as the RR-charge. Because of the
special properties of solutions in codimension 2, there is no
asymptotically flat region, except for the `surrogate asymptotic region'
around $r=1$. This means in particular that the standard supersymmetric
7-branes with RR-charge
$a_0>0$ is actually located at
$r=\infty$. When we wrap the 7-branes on a $K3$, we have in fact naked
singularities at each of the zero locus for $Z_7$ and $Z_3$
respectively, where $Z_3=0$ is the location of the repulson~\cite{joep}.
The \enh\ radius occurs at a radius where the effective
tension, $\mu^s_{3,7}\define(\mu_7 V^s_{K3}(r) - \mu_3)$ vanishes,
\ie\ where the volume
of the $K3$ reaches the value $\mu_3/\mu_7=(2\pi \sqrt{\alpha'})^4$
\eqn
     V^s(r) \equiv V^s_{K3} {Z_3(r) \over Z_7(r)} = {\mu_3\over \mu_7}~.
     \Label{e:V(r)SuSy}
\enn
Thus, the repulson singularity at $Z_3=0$ is resolved and the
branes which were located at the repulson singularity have now moved
{\it inwards} to the \enh\ radius implicitly given by the expression for
$V^s(r)$ above.

Returning to~(\ref{e:BIwrap}), we note the difference between the
supersymmetric expression~(\ref{e:V(r)SuSy}) and the corresponding result
in our case~(\ref{e:V(r)K3}). Hence, the action and
the tension for our 3-7 brane system, vanishes when
\begin{equation}
     \mu_{3,7}~\define~
               \mu_7V_{K3} \Big({Z_7(r)\over Z_3(r)}\Big)^{1/2}-\mu_3~=~0~.
\end{equation}
This happens at the location which we will call the \enh\ radius:
\begin{equation}
     \re ~\define~ e^{({V_*\over V_{K3}}-1)/a_0}~, \quad\hbox{where}\quad
     V_*~\define~(2\p\a'^{1/2})^4~.
\end{equation}
Note that in the $V_{K3}\to\infty$ limit, or alternatively $\a'\to 0$,
$\re\to \rs^+$.

Let us focus on the $a_0>0$ case; the $a_0<0$ case can then easily be
obtained by sending $r\to1/r$.
There appear three marked locations in the
physically permissible domain of $r$:
\begin{enumerate}
     \item $\rs^+=e^{-a_0^{-1}}$, where $Z_7(r)=0$, and which is the lower limit
           of the physically permissible region;
     \item $\re= e^{({V_*\over V_{K3}}-1)a_0^{-1}}\geq \rs^+$, where the
           effective tension, $\mu_{3,7}$ vanishes;
     \item $\rF=e^{{V_{K3}\over V_*}a_0^{-1}}\geq \re$, where $Z_3(r)=0$,
           and the dilaton diverges owing to Eq.~(\ref{e:DilZ3}).
\end{enumerate}
Clearly, near and beyond $\rF$, the stringy effects become too
strong for the present  analysis which needs to be further
refined; our present conclusions are therefore limited to $r<\rF$.

 From a naive supergravity point of view, the wrapped 7-branes are
located at the naked singularity, $\rs^+$. However, the brane probe does
not detect this singularity. Instead, the wrapped 7-branes have
effectively moved out from $\rs^+$ to $\re$.
Still, we have to make sure that the resolution of the naked singularity
takes place in a region in which the probe analysis is valid. We therefore
turn to this task.

\begin{figure}[ht]
       \framebox{\epsfxsize=160mm%
       \hfill~%Here be picture.
       \epsfbox{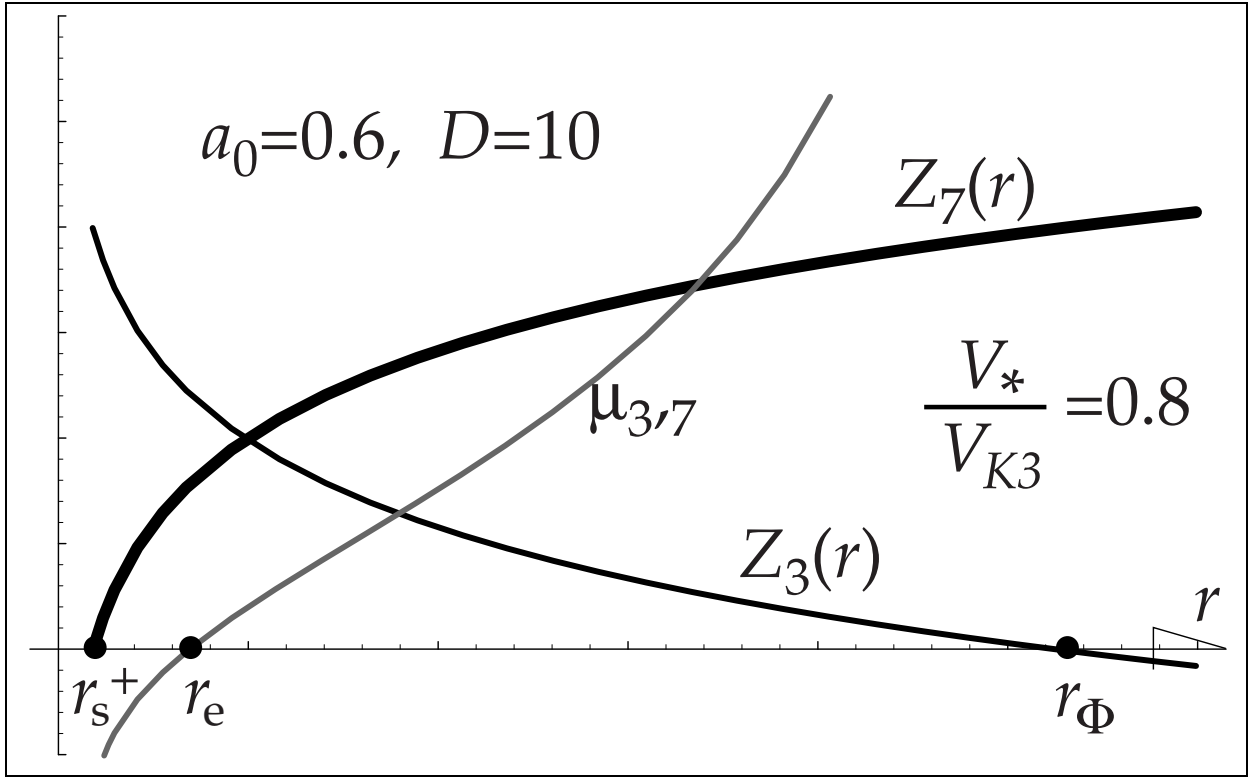}
       \hfill~}
       \caption{A sketch of the physically permissible region, with the
marked locations of $\rs^+$, $\re$ and $\rF$.}
       \Label{f:ThreeRs}
\end{figure}

Following the
analysis in section~\ref{s:PrT=const} we have a Lagrangian for the test
probe given by
\eqn
{\cal L} &=&
-g^{-1}_s Z_3 Z_7(\mu_7 V_{K3} Z_3^{-1/2} - \mu_3 Z_7^{-1/2}) +
\mu_7 V_{K3} C_8 - \mu_3 C_4 +\nn\ \\
&& {1\over 2} l^2 g^{-1}_s r^{-2} Z_3^{-1/8} Z_7^{-1/8}
(\mu_7 V_{K3} Z_3^{-1/2} - \mu_3 Z_7^{-1/2}) v^2~+~O(v^4)~.
\Label{e:L37free}
\enn

\begin{figure}[ht]
       \framebox{\epsfxsize=160mm%
       \hfill~%Here be picture.
       \epsfbox{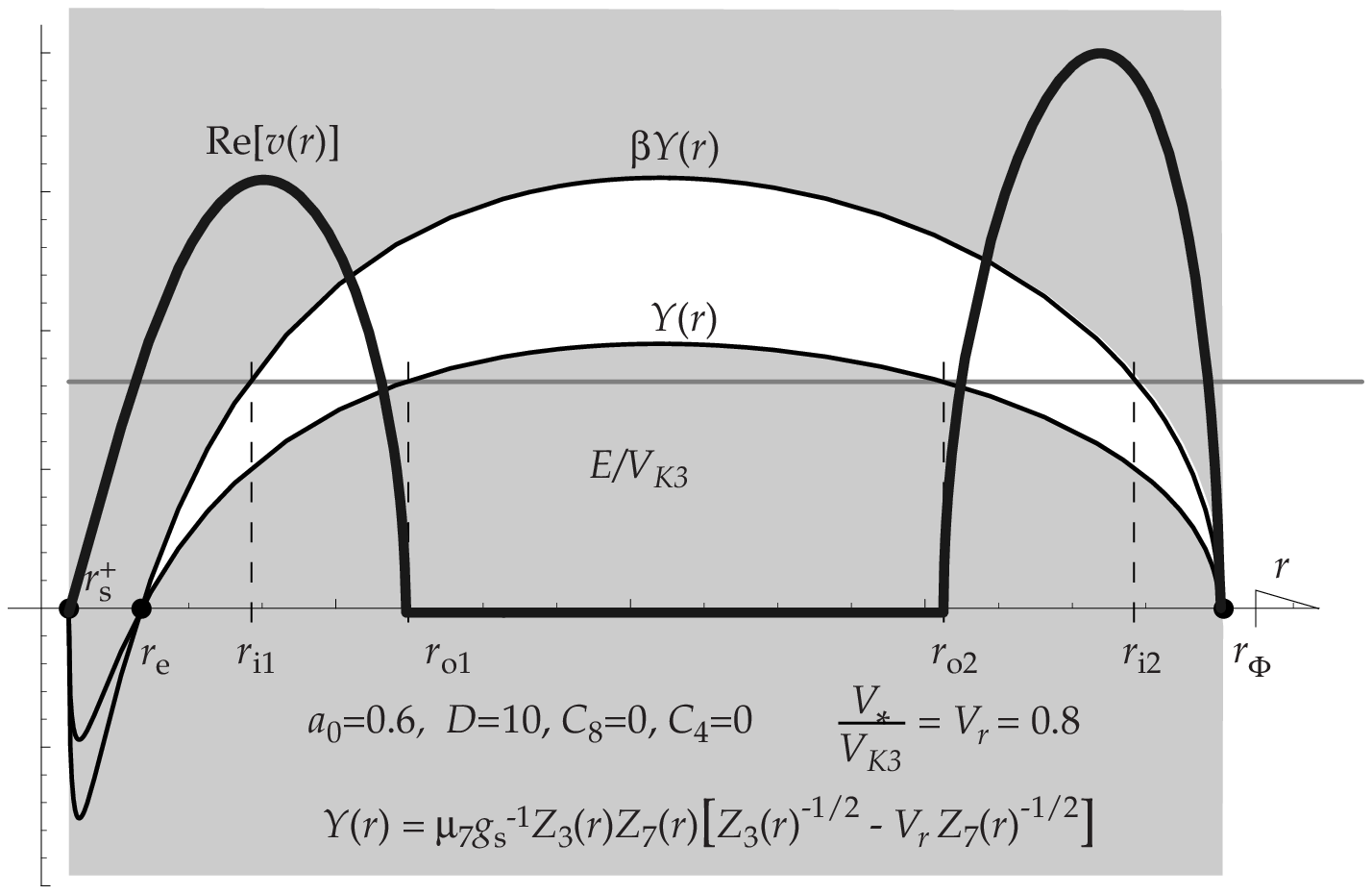}
       \hfill~}
       \caption{A sketch of the physically permissible region, with the
marked locations of $\rs^+$, $\re$, $r\mkern-1mu_{{\rm i}1,2}$,
$r\mkern-1mu_{{\rm o}1,2}$ and $\rF$.}
       \Label{f:LimSpd37}
\end{figure}

In a supersymmetric setting one finds that the effective potential for the
brane probe is constant. However, that is not the case above. As
in section~\ref{s:PrT=const} we assume that the total energy, $E$, is
conserved and find the following condition on the energy density $\cE$,
\begin{eqnarray}
      \mu_7[g_s^{-1}Z_3(r)Z_7(r)( Z_3^{-1/2} -
{V_*\over V_{K3}} Z_7^{-1/2}) - C_8  + C_4 {V_*\over V_{K3}}]&\leq&\cE~,\nn\\
\mu_7[\b g_s^{-1}Z_3(r)Z_7(r)(Z_3^{-1/2} -
{V_*\over V_{K3}} Z_7^{-1/2}) - C_8  + C_4 {V_*\over V_{K3}}]&\geq&\cE~.
     \Label{e:ExtendedK3}
\end{eqnarray}
These inequalities provide a region of allowed energies for
the probe to be at a given position, see Fig~\ref{f:LimSpd37}.
Note that compared to the unwrapped case, it is possible to have two
physically permissible regions,
$r\in[r\mkern-1mu_{{\rm i}1},r\mkern-1mu_{{\rm o}1}]$ and
$r\in[r\mkern-1mu_{{\rm o}2},r\mkern-1mu_{{\rm i}2}]$. We defer a more
detailed analysis of this point to~\cite{BHM2a}.

In the limit $a_0\to 0$, the inequalities~(\ref{e:ExtendedK3}) simplify
to
\begin{equation}
    \mu_7[g_s^{-1}(1-{V_*\over V_{K3}}) -  C_8 + C_4
{V_*\over V_{K3}}]~\leq~\cE~\leq~\mu_7[\b\,g_s^{-1}(1-{V_*\over V_{K3}}) -
C_8 + C_4 {V_*\over V_{K3}}]~.
\Label{e:SuSyWindowK3}
\end{equation}
 From Fig~\ref{f:LimSpd37} we find that the range of
validity is given by $\ri$ and $\ro$:
\begin{equation}
     \lim_{a_0\to0} r\mkern-1mu_{{\rm o}1}=+\infty~,\quad\hbox{and}\quad
     \lim_{a_0\to0} r\mkern-1mu_{{\rm i}1}=~0~=\lim_{a_0\to0} \rs^+~.
\end{equation}
That is, in the $a_0\to0$ limit, the probe analysis based on the
expansion~(\ref{e:BIactionzero}) is valid in the whole $r\in[0,\infty)$
region. Finally,  both the potential and kinetic terms
go to constants in~(\ref{e:L37free}) and
we are back in a supersymmetric scenario.

\subsection{The $T_{\mu\nu}\neq 0$ case}

\subsubsection{$\tau_I$}
This case resembles the source-free situation in that $\Ree\tau_I=\const$
and hence both $C_{0,8}$ are constant. Thus, the solution is given by the
following expressions,
\eqn
ds^2 &=& Z_3^{2/4} Z_7^{1/4}  \eta_{ab} dx^a dx^b +
V^{1/2}_{K3} Z_7^{1/4}ds^2_{K3} + \nn\ \\
&& l^2 Z_3^{-5/8} Z_3^{-7/8} e^{\xi a_0^{-1} - \xi a_0^{-1} Z_7^2}
e^{\xi a_0^{-1} - \xi a_0^{-1} Z_3^2}
r^{-2}(dr^2 + r^2 d\theta^2)\\
e^{2\F} &=& g_s^2 e^{-2\omega \theta} Z_3^{-1}\\
C_4&=&\const,\quad
C_8=\const,\\
Z_3&=& 1 - a_0 {(2\pi \a'^{1/2})^4\over V_{K3}} \log(r)\\
Z_7&=& 1 + a_0  \log(r)~.
\enn

Although there is a non-trivial $\theta$ dependence in the dilaton
which gives a slight modification to the metric the
probe analysis goes through as in section~5.1,
\eqn
{\cal L} &\approx&
-g_s^{-1}e^{-\w\q}Z_3^{1/2} Z_7^{1/2}(\mu_7 V_{K3} Z_7^{1/2} - \mu_3
Z_3^{1/2}) +
\mu_7 V_{K3} C_8 - \mu_3 C_4+\nn \\
&&
{1\over 2} l^2 g_s^{-1} e^{\omega \theta}
(\prod_{i=3,7}e^{\xi a_0^{-1} - \xi a_0^{-1} Z_i^2}Z_i^{-9/8})
r^{-2}
(\mu_7 V_{K3} Z_7^{1/2} - \mu_3 Z_3^{1/2}) v^2~.
\Label{e:L37tau1}
\enn

%%%The analysis goes through as in the source-free case.
As in section~\ref{s:enhancon} we assume that the total energy, $E$, is
conserved and find the following condition on the energy density $\cE$,
\begin{eqnarray}
      \mu_7[g_s^{-1}e^{-\w\q}Z_3(r)Z_7(r)(Z_3^{-1/2} -
{V_*\over V_{K3}} Z_7^{-1/2}) - C_8 + C_4 {V_*\over V_{K3}}]&\leq&\cE~,\nn\\
  \mu_7[\b\,g_s^{-1}e^{-\w\q}Z_3(r)Z_7(r)(Z_3^{-1/2} -
{V_*\over V_{K3}} Z_7^{-1/2}) - C_8 + C_4 {V_*\over V_{K3}}]&\geq&\cE
~.
     \Label{e:ExtendedK3I}
\end{eqnarray}
These inequalities provide a region of allowed energies for
the probe to be at a given position.
For any given direction, $\q$, in the transverse space we can sketch
the physical permissible region as in~Fig.\ref{f:LimSpd37}.

In the limit $a_0\to 0$, the inequalities~(\ref{e:ExtendedK3I}) simplify
to
\begin{equation}
    \mu_7[g_s^{-1}(1-{V_*\over V_{K3}}) -  C_8 + C_4
{V_*\over V_{K3}}]~\leq~\cE~\leq~\mu_7[\b\,g_s^{-1}(1-{V_*\over
V_{K3}}) -  C_8 + C_4 {V_*\over V_{K3}}]~.
\Label{e:SuSyWindowK3I}
\end{equation}
Following the same procedure as in the previous section, we find the
following range of validity:
\begin{equation}
     \lim_{a_0\to0} \ro=+\infty~,\quad\hbox{and}\quad
     \lim_{a_0\to0} \ri=~0~=\lim_{a_0\to0} \rs^+~.
\end{equation}
That is, in the $a_0\to0$ limit, the probe analysis based on the
expansion~(\ref{e:BIactionzero}) is valid in the whole $r\in[0,\infty)$
region.

Finally, in the limit $a_0\to 0$ the
potential term in~(\ref{e:L37tau1}) becomes a constant. However, the
kinetic energy has an $r$-dependence from the non-trivial dilaton,
$r^{\xi(-1 + (2\pi\sqrt{\alpha'})^4/V_{K3})}$, in addition to the
cylindrical $r^{-2}$ factor. The exponent reflects the deficit angle
created by the wrapped 7-brane. How does this compare to the
supersymmetric case? A collection of supersymmetric 7-branes located
at the $r=\infty$ has the same $r^{-\xi}$ behavior in the metric to
reflect a deficit angle $\Delta=2\pi|\xi|$, \ie\ the transverse space
is a cone rather than a plane. By redefining the coordinates on the
transverse space, we can thus absorb this conical behavior in the
induced metric on the brane probe, and hence in the velocity of the
brane probe. This argument can be repeated for the wrapped case.
In the end, the effective mass of the probe is constant.

\subsubsection{$\tau_{II}$}

Based on our earlier arguments the wrapped $\tau_{II}$ solution is
given by
\eqn
   ds^2 &=& Z_3^{2/4} Z_7^{1/4}  \eta_{ab} dx^a dx^b +
   V^{1/2}_{K3} Z_7^{1/4}ds^2_{K3} + \nn\ \\
   && l^2 Z_3^{-5/8} Z_3^{-7/8} e^{\xi a_0^{-1} - \xi a_0^{-1} Z_7^2}
   e^{\xi a_0^{-1} - \xi a_0^{-1} Z_3^2}
   r^{-2}(dr^2 + r^2 d\theta^2)\\
   e^{2\F} &=& g_s^{2} f(\q)^2 Z_3^{-1}~,
   \quad\hbox{where}\quad f(\q)\define2\tanh(\w\pi)\cosh(\w\q)\\
   C_4&=&{\omega\over a_0} \tanh(\omega\pi) Z_3^2(r)\\
      C_8&=&{\omega\over a_0} \tanh(\omega\pi) Z_7^2(r)\\
   Z_3&=& 1 - a_0 {(2\pi \a'^{1/2})^4\over V_{K3}} \log(r)\\
   Z_7&=& 1 + a_0  \log(r)~.
\enn

Since the RR-potentials that couple to the 3 and 7-branes are non-trivial
the effective Lagrangian as seen by the the brane probe is given by (to
$O(v^2)$)
\begin{eqnarray}
{\cal L} &=& -g_s^{-1} f(\q)
    Z_3^{1/2} Z_7^{1/2}(\mu_7 V_{K3} Z_7^{1/2} - \mu_3 Z_3^{1/2}) +\nn\\
   &&{\omega\over a_0} \tanh(\w\pi) (\mu_7 V_{K3}Z_7^2(r)  C_8^{(0)} -
    \mu_3 Z_3^2(r)  C_4^{(0)}) + \\
   && {1\over 2} l^2 g^{-1}_s f(\q)
      (\prod_{i=3,7}e^{\xi a_0^{-1} - \xi a_0^{-1} Z_i^2} Z_i^{-9/8}) r^{-2}
      (\mu_7 V_{K3} Z_7^{1/2} - \mu_3 Z_3^{1/2}) v^2~.
   \Label{e:L37tau2}
\end{eqnarray}

The analysis goes through as above ,
\begin{eqnarray}
    \mu_7[g_s^{-1}f(\q)Z_3(r)Z_7(r)(
       Z_3^{-1/2} - {V_*\over V_{K3}} Z_7^{-1/2}) - C_8 + C_4 {V_*\over V_{K3}}]
       &\leq&\cE~,\nn\\
    \mu_7[\b\,g_s^{-1}f(\q)Z_3(r)Z_7(r)(
      Z_3^{-1/2} - {V_*\over V_{K3}} Z_7^{-1/2}) - C_8 + C_4
{V_*\over V_{K3}}]&\geq&\cE~.
     \Label{e:ExtendedK3II}
\end{eqnarray}
These inequalities provide a region of allowed energies for
the probe to be at a given position.
As was noted in section~3.2.2, in the $a_0\to 0$ limit
the probe analysis breaks down.

\subsection{Exponential hierarchy}
\Label{s:exphierarchy}
Finally let us compute the ratio between the $D$ and $D-2$ dimensional
Planck scales following the procedure outlined in our previous
paper~\cite{bhmone,cohen}. In particular, the relation between the eight
and ten dimensional scales is (using again $\rho=a_0^{-1}[1+a_0\log(r)]$)
\begin{equation}
M_8^6~=~M_{10}^8 \int e^{6A+2B}\rd\rho\rd\q~=~
\p l^2 M_{10}^8\, e^{\xi/a_0}\rho_o^{9/16}\gamma({7\over 16},\rho_o)~,
\label{e:volume}
\end{equation}
where $\rho_o(\ro)$ represents the upper
limit of the {\it physically\/} allowed region $r\in[\rs^+,\ro]$;
the integral vanishes in the lower limit, $\rs^+$.
   Clearly, the exponential hierarchy driven by $\xi/a_0$ remains
essentially unchanged.

There is a formula analogous to~(\ref{e:volume}) also in the case when our
background is wrapped on a $K3$. However, depending on the energy of the
probe, there are two possible cases; see Fig.~\ref{f:LimSpd37}. If $\cE$
is larger than the maximum value of the lower limit function in the
inequality~(\ref{e:ExtendedK3}), then $\re\leq r\leq r_\F$, and the
analogue of Eq.~(\ref{e:volume}) has a single contribution.
 On the other hand, if $\cE$ is less than that, then the probe is allowed
in two disconnected regions: $\re\leq r\leq r_{{\rm o}1}$ and
$r_{{\rm o}2}\leq r\leq r_\F$. All of this will however only modify the
value of the coefficient in the analogue of Eq.~(\ref{e:volume}); the
exponential hierarchy remains governed by $e^{\xi/a_0}$, as derived in
Ref.~\cite{cohen,bhmone}.

\section{Conclusions}

In this paper we have presented a detailed analysis of the space-time
properties of a certain family of non-supersymmetric vacua which
can lead to an exponential hierarchy for the string coupling of
$O(1)$. In particular we have identified a limit in which
these non-supersymmetric vacua can be interpreted from the point
of view of F-theory. In principle, the dynamical breaking of supersymmetry
of these F-theory models of particle physics should be tied to
the emergence of exponential hierarchy.
It would be very interesting to understand this more precisely.

The exponential hierarchy is due to the presence of a classical naked
singularity of our background. We have shown that this singularity
can be resolved via a non-supersymmetric effect analogous to the \enh\
mechanism, which in turn is based on a repulsive nature of the
naked singularity as seen by quantum probes.

One potentially very exciting feature of our solution is that
the constant $\xi/a_0$ which appears in the expression for the ratio
of the $D$ and $D-2$ dimensional Planck scales is undetermined and
therefore is an input parameter.
Thus, in principle one might wonder if we can have
arbitrarily large exponential hierarchy. (This is similar to some recent
discussion on the cosmological constant problem from a ``holographic''
point of view~\cite{cc}.) In our case, the emergence of exponential
hierarchy might be tied to a dynamical supersymmetry breaking of a
particular class of F-theory vacua. It is natural to expect that
the value of $\xi/a_0$ is
related to the  supersymmetry breaking scale.

Finally, one
aspect of classical gravitation analysis not addressed in this
paper is the issue of classical stability.
Recall that the general perturbation of the metric $g_{\mu \nu} \to g_{\mu \nu}
+ h_{\mu \nu}$ satisfies (in a suitable gauge) the Lichnerowicz
equation \cite{stability}
\begin{equation}
\Delta_{L} h_{\mu \nu} = [\delta_{\mu}^{\mu_1} \delta_{\nu}^{\nu_1} \Box
+2 {R_{\mu}^{\mu_1}}_{\nu}^{\nu_1}] h_{\mu_1 \nu_1} = 0
\end{equation}
The general perturbation is split into scalar, vector and tensor
modes and the stability analysis is reduced to finding whether there exist
any unstable modes for the perturbation.
Work along these lines is in progress \cite{andrew}.

{\bf Acknowledgments:}
We thank V.~Balasubramanian, O.~Bergman, P.~Candelas,
A.~Chamblin, R.~Corrado, M.~Cveti\v{c}, J.~de~Boer, J.~Gomis, R.~Gregory,
A.~Karch, B.~Kol,
C.~Johnson, R.~Leigh, D.~Marolf, J.~Rahmfeld, M.~Strassler, H.~Verlinde
and N. Warner for useful discussions.
       The work of P.~B.\ was supported in part by  the US
Department of Energy under grant number DE-FG03-84ER40168.
P.B. would like to thank Spinoza Institute and Oxford University
for their hospitality in the formative stages of this project.
       T.~H.\ wishes to thank the US Department of Energy for their
generous support under grant number DE-FG02-94ER-40854 and the
Institute for Theoretical Physics at Santa Barbara, where part of
this work was done with the support from the National Science
Foundation, under the Grant No.~PHY94-07194.
       The work of D.~M.\ was supported in part by the US
Department of Energy under grant number DE-FG03-84ER40168.
D.~M.\ would like to thank Howard University, the University of
Illinois at Chicago, the University of Illinois at
Urbana and the University of Pennsylvania
for their hospitality while this work was in
progress.

\end{document}